\documentclass{aa}
\usepackage{natbib,psfig,graphicx,ulem}
\usepackage{txfonts}
\usepackage{longtable}
\usepackage{soul,color}

\def\mathnew{\mathsurround=0pt}
\def\simov#1#2{\lower .5pt\vbox{\baselineskip0pt \lineskip-.5pt
\ialign{$\mathnew#1\hfil##\hfil$\crcr#2\crcr\sim\crcr}}}

\def\MeV{Me\kern-0.11em V}
\def\keV{ke\kern-0.11em V}

\begin{document}

\title{Comparison of the properties of two fossil groups of galaxies with the
normal group NGC~6034 based on multiband
  imaging and optical spectroscopy \thanks{Based on observations made
    at Observatoire de Haute Provence (CNRS), France, at Asiago
    Observatory (Italy), at TNG (La Palma, Spain), and with ESO
    Telescopes at the La Silla Observatory under programme ID
    082.A-0374.  Also based on the use of the NASA/IPAC Extragalactic
    Database (NED) which is operated by the Jet Propulsion Laboratory,
    California Institute of Technology, under contract with the
    National Aeronautics and Space Administration, and on the use of
    the IA2-TNG archive (constructed as part of the activities of the
    Italian Theoretical Virtual Observatory).  Funding for the SDSS
    and SDSS-II has been provided by the Alfred P. Sloan Foundation,
    the Participating Institutions, the National Science Foundation,
    the U.S. Department of Energy, the National Aeronautics and Space
    Administration, the Japanese Monbukagakusho, the Max Planck
    Society, and the Higher Education Funding Council for England. The
    SDSS Web Site is http://www.sdss.org/.  The SDSS is managed by the
    Astrophysical Research Consortium for the Participating
    Institutions. The Participating Institutions are the American
    Museum of Natural History, Astrophysical Institute Potsdam,
    University of Basel, University of Cambridge, Case Western Reserve
    University, University of Chicago, Drexel University, Fermilab,
    the Institute for Advanced Study, the Japan Participation Group,
    Johns Hopkins University, the Joint Institute for Nuclear
    Astrophysics, the Kavli Institute for Particle Astrophysics and
    Cosmology, the Korean Scientist Group, the Chinese Academy of
    Sciences (LAMOST), Los Alamos National Laboratory, the
    Max-Planck-Institute for Astronomy (MPIA), the
    Max-Planck-Institute for Astrophysics (MPA), New Mexico State
    University, Ohio State University, University of Pittsburgh,
    University of Portsmouth, Princeton University, the United States
    Naval Observatory, and the University of Washington.  }}

\author{C.~Adami\inst{1} \and 
S.~Jouvel\inst{1,2} \and
L.~Guennou\inst{1} \and 
V.~Le Brun\inst{1} \and 
F.~Durret\inst{3,4} \and 
B.~Clement\inst{1} \and
N.~Clerc\inst{5} \and
S.~Comer\'on\inst{6} \and 
O.~Ilbert\inst{1} \and
Y.~Lin\inst{7} \and 
D.~Russeil\inst{1}  \and 
U.~Seemann\inst{8,9} 
}

\offprints{C. Adami \email{christophe.adami@oamp.fr}}

\institute{
LAM, OAMP, Universit\'e Aix-Marseille $\&$ CNRS, P\^ole de l'Etoile, Site de
Ch\^ateau Gombert, 38 rue Fr\'ed\'eric Joliot-Curie,
13388 Marseille~13 Cedex, France
\and
UCL, Department of Physics \& Astronomy, Gower Place,
London WC1E 6BT, United Kingdom
\and 
UPMC Universit\'e Paris 06, UMR~7095, Institut d'Astrophysique de Paris,
98bis Bd Arago, 75014, Paris, France
\and
CNRS, UMR~7095, Institut d'Astrophysique de Paris, 75014, Paris, France
\and
Laboratoire AIM, CEA/DSM/IRFU/Sap, CEA-Saclay, 91191 Gif-sur-Yvette Cedex, France
\and
Korea Astronomy and Space Science Institute, 776, Daedeokdae-ro, Yuseong-gu, 
Daejeon, 305-348 Republic of Korea 
\and
Centre de Recherche Astrophysique de Lyon, Universit\'e de Lyon, Universit\'e Lyon 1, Observatoire de Lyon, Ecole Normale Sup\'erieure de Lyon, CNRS, France
\and
Institut fur Astrophysik, Georg-August Universitat Gottingen, Friedrich-Hund-Platz 1, 37077 Gottingen, Germany
\and
European Southern Observatory, Karl-Schwarzschild-Str. 2, 85748 Garching, Germany
}

\date{Accepted . Received ; Draft printed: \today}

\authorrunning{Adami et al.}

\titlerunning{Comparison of the properties of two fossil groups of galaxies 
with the normal group NGC~6034}

\abstract 
{Fossil groups are dominated by a bright galaxy, and their luminosity
functions show an absence within half the virial radius of galaxies
brighter than the central galaxy magnitude +2. They are nevertheless
massive with an extended X-ray halo. The formation and evolution of
these structures is still widely debated.}
{To better understand the origin of these structures, it is
  crucial to study their faint galaxy population, as well as their
  large-scale environment, to determine in particular whether they are
  isolated or not.}
{We collected multiband imaging and spectroscopy for two fossil groups
  (RX J1119.7+2126 and 1RXS J235814.4+150524) and one normal group
  (associated with NGC~6034). We computed photometric redshifts in the
  central zones of each group, combining previous data with the SDSS
  five-band data. For each group we investigated the red sequence (RS) of the
  color-magnitude relation and computed the luminosity functions,
  stellar population ages and distributions of the group
  members. Spectroscopy allowed us to investigate the large-scale
  surroundings of these groups and the substructure levels in 1RXS
  J235814.4+150524 and NGC~6034.}
{ The large-scale environment of 1RXS J235814.4+150524 is poor, though
  its galaxy density map shows a clear signature of the surrounding
  cosmic web.  RX J1119.7+2126 appears to be very isolated, while the cosmic
  environment of NGC~6034 is very rich. At the group scale, 1RXS
  J235814.4+150524 shows no substructure. Galaxies with recent stellar
  populations seem preferentially located in the group outskirts. A
  red sequence is discernable for all three groups in a color-magnitude
  diagram. The luminosity functions based on photometric redshift
  selection and on statistical background subtraction have comparable
  shapes, and agree with the few points obtained from spectroscopic
  redshifts. These luminosity functions show the expected dip between
  first and second brightest galaxies for the fossil groups only.
  Their shape is also regular and relatively flat at faint magnitudes down 
  to the completeness level for
  RX J1119.7+2126 and NGC~6034, while there is a clear lack of faint galaxies
  for 1RXS J235814.4+150524. The faint parts of the
  luminosity functions appear dominantly populated by late-type
  galaxies. }
{RX J1119.7+2126 is definitely classified as a fossil group; 1RXS
  J235814.4+150524 also has properties very close to those of a fossil
  group, while we confirm that NGC~6034 is a normal group.  }


\maketitle

\section{Introduction}

The galaxy population of a 
fossil group is dominated by a giant elliptical (D) galaxy and includes in 
addition only galaxies at least two magnitudes fainter than D following the 
definition of Jones et al. (2003, see also Proctor et al. 2011 for a less 
stringent definition). These structures
have been discovered only quite recently in the late 90s,
concomitantly with the arrival of large-scale X-ray surveys (see
e.g. Adami et al. 2011 for a recent catalogue of X-ray structures
including a group of galaxies very close to the fossil group
status). This is because in the optical only, fossil group X-ray halos are 
not a-priori detectable. The galaxy population of fossil groups is quite 
different from normal galaxy groups of similar mass, at least as far as 
the bright end of the galaxy luminosity function is concerned (because of 
the lack 
of galaxies in the first two magnitude bins of the luminosity functions)
and the origin of these structures is still widely debated.
 They could result from originally truncated galaxy
luminosity functions (Mulchaey $\&$ Zabludoff 1999), or be galaxy
structures that have been depopulated in terms of galaxies by merging
processes (e.g. Cypriano et al. 2006). We also refer the reader to the recent paper 
by Proctor et al. (2011), which could lead to a new vision of the fossil systems formation
and evolution.

To better understand the origin of these fossil structures,
it is therefore crucial to study their faint galaxy populations. This
is a demanding task because even for nearby objects the galaxies
considered are often fainter than Rc$\sim$20. This requires large
telescopes to achieve spectroscopy (e.g. Lopes de Oliveira et al. 2010
and references therein), though deep imaging can be performed with smaller
telescopes (e.g. Adami et al. 2007b). It is also important to study
the large-scale environment of these structures to determine whether they
are isolated or not. This is also a difficult task because it requires
spectroscopic samples of bright galaxies on typical scales of
$\sim$100 deg$^2$. The very low surface density of these targets makes
the large multi-spectrographs quite inefficient and we can use instead
single-slit instruments on small telescopes to complete the bright
spectroscopic catalogs already available, such as the Sloan Digitalized Sky
Survey (SDSS).

We chose a triple approach, using small telescopes (2m
class) to obtain very large field spectroscopic samples of bright
galaxies in the group surroundings, and direct, relatively deep imaging
of the groups themselves (typically in one virial radius), and larger
telescopes (4m class) to obtain deep spectroscopy of faint group galaxy
members. Our targets are two fossil groups (RX J1119.7+2126 and 1RXS
J235814.4+150524) and one normal group (associated with NGC~6034, which
we will call in the following NGC~6034 for simplicity). These data
will allow us to compare the behaviors of these two different
classes of structures in a homogeneous way.

Section 2 describes our data sets. Section 3 deals with the nature of
the considered groups. Section 4 describes the large-scale environment
of 1RXS J235814.4+150524.  Section 5 gives details on the small-scale
behavior of the groups (including color-magnitude relations and
galaxy luminosity functions). Finally Section 6 is the summary.

Throughout the paper we assume H$_0$ = 71 km s$^{-1}$ Mpc$^{-1}$,
$\Omega _m$=0.27, and $\Omega _{\Lambda}$=0.73 (Dunkley et al.
2009). All magnitudes are given in the AB system.

\section{Sample, observations, and data characteristics}

We selected twostructures among the known fossil groups (FG hereafter).  
We already dedicated an article to the first one (Adami
et al. 2007b on RX J1119.7+2126.7) and we present additional extensive
imaging in the present work. The second group (1RXS J235814.4+150524)
was selected to be one of the most distant known FGs (z$\sim$0.17,
Santos et al. 2007, but see also the z$\sim$0.6 structure of Ulmer et
al. 2005, close to the fossil group status).  The third selected
target is the group of galaxies associated with NGC~6034. It was
originally presented as a FG by Yoshioka et al. (2004), but Lopes de
Oliveira et al. (2010) showed that this group was in fact a normal
group of galaxies, and we confirm this in the present work. This
structure is very nearby, so we can sample its galaxy populations down
to relatively deep absolute magnitudes even with small
telescopes. These three groups are sampled by SDSS photometry and
spectroscopic redshifts. We give their main
characteristics in Table~\ref{tab:caract} .

\begin{table*}
  \caption{Main characteristics of the three studied groups: name,
    coordinates, redshift, bolometric X-ray luminosity 
    (in units of 10$^{42}$ erg~s$^{-1}$) from
    Lopes de Oliveira (2010), Jones et al. (2003), and Santos et al. (2007) 
    respectively, converted to the presently adopted cosmology and to 
    bolometric values when needed using web PIMMS, optical R-band total luminosity, velocity dispersion, virial 
    radius, and covered area (in percentage of the virial radius) by the 
    imaging data.}
\begin{center}
\begin{tabular}{cccccccccc}
\hline
\hline
Name  & RA & DEC & z & L$_{X,bol}$ & R lum. & Vel. disp. & virial rad. & Radial sampling \\
      & (J2000.0)  & (J2000.0)  &          & 10$^{42}$ erg/s & 10$^{11}$ L$\odot$  & km/s & Mpc & $\%$ virial rad. \\
\hline
RX J1119.7+2126        & 11h19min43.7s & +21deg26'50'' & 0.061 & ~0.84 & 0.8 & 294 & 0.72 & 56$\%$\\ 
1RXS J235814.4+150524  & 23h58min14.4s & +15deg05'24'' & 0.178 & 14.02 & 3.2 & 254/578 & 0.58/1.31 & $\ge$76$\%$\\ 
NGC~6034               & 16h03min32.1s & +17deg11'55'' & 0.034 & 13.88 & 3.1 & 297 & 0.71 & 56$\%$\\ 
\hline
\end{tabular}
\label{tab:caract}
\end{center}
\end{table*}

The main goals of the present work are to study their faint galaxy
populations through spectroscopy and photometric redshifts and to
sample the large-scale environment of these groups. We therefore
collected PI observations completed by archive data as described in
the following.

\subsection{Spectroscopic data}

We first compiled/reduced public Nasa Extragalactic Database (NED) and 
TNG (Telescopio Nazionale Galileo) spectroscopic data in the
regions defined below for the three groups of the sample, and then we
performed additional spectroscopy with various telescopes.

\subsubsection{Spectroscopy of bright galaxies}

As already mentioned, the isolation level of a fossil group from the
cosmic web is a key element to constrain its origin. We have shown in
Adami et al. (2007b) that given the field luminosity function at low
redshift, the probability is low to have a galaxy brighter than the dominant
fossil group galaxy in the surrounding cosmic web portions unless 
a massive galaxy structure is present in the region. In other words, if 
there is no galaxy
brighter than the fossil group dominant galaxy in a given cosmological
volume, the fossil group is likely to be the dominant structure of
this volume. Since fossil groups are relatively minor structures, 
if they are the dominant structures of a given volume,
this volume is also likely to be relatively poorly populated. This
kind of study was already conducted by Adami et al. (2007b) for RX
J1119.7+2126 and for NGC~6034 by Lopes de Oliveira (2010).

In the present work we searched for galaxies brighter than the
dominant galaxy of 1RXS J235814.4+150524 (r'=16.3) in a
5$\times$5~deg$^2$ area around the group. This roughly corresponds to
a 50$\times$50~Mpc$^2$ area, which agrees well with the maximum size
of the known voids (Hoyle $\&$ Vogeley 2004). To perform this
search, we selected all objects classified as galaxies
in the quoted area from the SDSS that are brighter than r'=16.3, and do not 
have a known
spectroscopic redshift. We visually inspected all selected objects
to remove obvious saturated stars and finally had a list of
18 galaxies. Given the magnitudes of these objects, we used 2m class
telescopes to observe them spectroscopically. Specifically, we made
observations with the CARELEC instrument mounted on the 1.93m
telescope at OHP in 2007 and 2008, and with the Boller $\&$ Chivens
and AFOSC instruments mounted on the 1.22m and 1.8m telescopes of the
Asiago observatory in 2007. We successfully measured redshifts for 16
of the 18 targets and not for all of them because of time limitations. 
The exposure times and new 
redshifts acquired are listed in Table~\ref{tab:spectroohp}, and we show
an example of these 16 spectra in the appendix (see Fig.~\ref{fig:J23480439}).
Data reductions were made with the MIDAS-based tool that was already 
described in 
Adami et al. (1998). Redshifts were measured with the EZ tool (Garilli et al. 
2010) and dedicated MIDAS routines. 

\subsubsection{Spectroscopy of faint galaxies}

When considering fossil groups, the two
magnitude gap between the brightest and the second brightest galaxy
makes any spectroscopic follow up of the faint galaxy
populations difficult. 2m class telescopes are therefore inefficient for this
task, which is better achievable with classical multiobject
spectrographs mounted on 4m class telescopes. Their relatively small
field of view is also well-adapted to the typical size of groups in
the redshift range of interest, typically a few arcminutes on the sky.

We observed 1RXS J235814.4+150524 and RX J1119.7+2126 using the NTT/EFOSC2 
(ID 082.A-0374) and the TNG/Dolores (2008B: CAT-9) spectrographs
in 2008. The details of the observations are given in
Table~\ref{tab:spectrodeep}. The same table also gives details about
the data that we collected from the TNG archive facility and reduced
for 1RXS J235814.4+150524 and RX J1119.7+2126.

\begin{table*}
\begin{center}
  \caption{Spectroscopic observations. Col.~1: group name; col.~2: instrument; 
    col.~3:  exposure time per mask; col.~4: grism; col.~5: allocation number. 
    The two first lines are our own observations. 
    The following lines are archive data.}
\begin{tabular}{lllll}
\hline
\hline
Name & Instrument & Exp. time & Grism & allocation \\ 
 &  & (s) & &  \\ 
\hline
1RXS J235814.4+150524 & ESO/NTT/EFOSC2 & 2400 & GR6 & 082.A-0374 \\
RX J1119.7+2126 & TNG/Dolores & 2400 & LR-R & A18 CAT-9\\
\hline
1RXS J235814.4+150524 & TNG/Dolores & 3600 & LR-B & A18 TAC-27 \\
RX J1119.7+2126 & TNG/Dolores & 1350/5400/7200 & MR-B & A11 TAC-27 \\
\hline
\end{tabular}
\label{tab:spectrodeep}
\end{center}
\end{table*}

The NTT and TNG data reductions were made with the MIDAS-based tool
that was already described in Adami et al. (1998). Combining the resulting
redshifts with those from the NED database, we ended up with catalogs of
43 and 32 redshifts for the 1RXS J235814.4+150524 and RX J1119.7+2126
lines of sight, respectively.

\begin{figure}
\centering \mbox{\psfig{figure=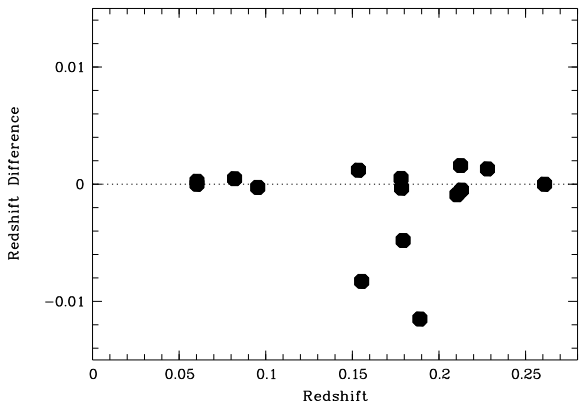,width=8cm,angle=270}}
\caption[]{Redshift difference between multiple redshift measurements
as a function of redshift. The horizontal dotted line represents the zero
difference.}
\label{fig:comparz}
\end{figure}

Given the sometimes limited signal-to-noise ratio of the A11 TAC-27 spectra 
 (S/N$\leq$3), we were forced to smooth the data for the considered galaxies. 
We note that redshifts were measured with the EZ tool (Garilli et al. 
2010) and dedicated MIDAS routines. To assess
the quality of our redshift measurements, we also checked that
multiple measurements of the same galaxies were giving similar
results. Merging the galaxy sample spectroscopically measured several
times in the PI spectroscopic runs (4 galaxies) and the remeasured
literature galaxies with a known spectroscopic redshift (11 galaxies),
we produced Fig.~\ref{fig:comparz}. Despite the low numbers involved, 
it appears that our measurements are quite secure with the exception of
one galaxy that presents two clearly discrepant redshift measurements (not shown
in Fig.~\ref{fig:comparz}).  This galaxy
is found at a 0.4937 redshift in the TNG data, while literature gives a value of
0.4460. The difference arises because the literature value is placing 
different lines in real absorption features compared to our measurement. 
In the literature the presence of H$\&$K, H$\delta$, and H$\beta$ was only
assumed, while we are
able to fit H$\&$K, H$\delta$, G band, H$\gamma$, and H$\beta$ in real 
absorption features. We are therefore assuming our value as the real redshift.

Three other galaxies (identified in Table~\ref{tab:spectrofaint2359}) also 
present slightly different redshift measurements in Fig.~\ref{fig:comparz} 
that are on the order of 0.01. 

- For the first one (z=0.1471 or 0.15539), the literature
value is mainly based on the NaD line, and only offers a poor fit of the H$\&$K 
lines. Our spectrum unambiguous shows these H$\&$K lines and leads to a redshift
of 0.1471. 

- The second one (z=0.1775 or 0.18900) offers two possibilities to fit the 
H$\&$K lines. Because the NTT spectrum has a clearly poorer S/N, we preferred
the redshift value based on the TNG spectrum (z=0.1775).

- The third one (z=0.1745 or 0.1793) is also based on a low S/N NTT spectrum 
and we preferred the TNG-based value of 0.1745. The difference in redshift 
arises because we may fitted the G band and H$\gamma$ instead of H$\&$K into 
a real double absorption line feature.

In conclusion we computed a 1$\sigma$ mean uncertainty of 0.0008 when 
excluding the four worst cases.

\subsection{Optical imaging}

Two of our main goals are to constrain the luminosity function and the
color-magnitude red sequence of the groups. To do this, we determined a 
membership criterion for faint galaxies. Our
spectroscopic catalogs are far from complete at these magnitudes and
we have to use photometric techniques such as photometric redshifts,
color-magnitude relations or field statistical comparisons. The first
step toward this goal is to obtain sufficiently deep photometric catalogs
for the considered groups.

\begin{figure}
\centering \mbox{\psfig{figure=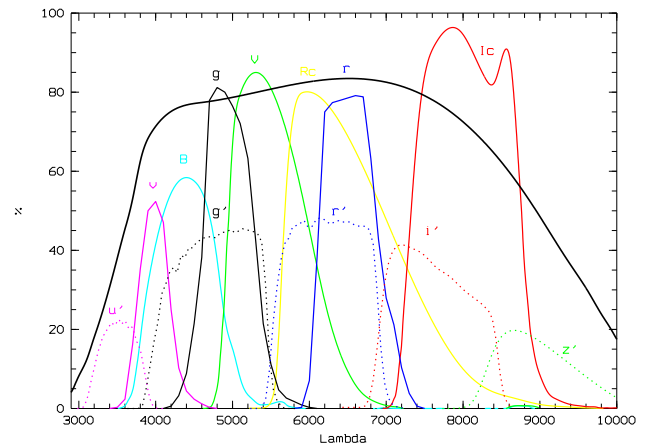,width=9cm,angle=270}}
\caption[]{Response functions of the used filters. The continuous thick 
black line gives the OHP T120 CCD
  response function. SDSS filter response functions (dotted curves) are given
  in arbitrary units.}
\label{fig:filt}
\end{figure}

In this framework and to complement the u', g', r', i', and
z' SDSS images, we obtained images of 1RXS J235814.4+150524 (a single field 
in the Rc filter) and NGC~6034 (a four field mosaic in the B and Rc filters) 
with the 1.2m OHP telescope. We also obtained multiband single-field images 
with the same telescope (OHP v, B, V, g, Rc, r, Ic filters,
see http://www.obs-hp.fr/guide/camera-120/ubvri.html, Cousins 1973, 1974, 
Thuan $\&$ Gunn 1976, and Schneider et al. 1983) with similar
depth for RX J1119.7+2126 (B and Rc-band data were already presented
in Adami et al.  2007b). Images were taken with the T120 1024x1024 CCD
camera and were observed under typical seeings between 2.5 and 3
arcsec. Details of the observations and of the SDSS images are given
in Table~\ref{tab:imaging}. We also show in Fig.~\ref{fig:filt} the
response functions of the selected filters.

\begin{table*}
  \caption{Imaging observations. Col.~1: group name, col.~2: band, col.~3: 
    total duration of the summed exposures (when OHP data), col.~4: number of 
    different fields on the 
    same target (when OHP data), col.~5: seeing, col.~6:  zero point 
    shift when computing photometric redshifts.}
\begin{tabular}{lllllr}
\hline
\hline
Name & Filter & Exposure time & Number of fields & Seeing & ZP shift \\ 
 &  & (s) &  & (arcsec) & (mag) \\ 
\hline
NGC~6034 & u' & - & - & 1.4 & -0.15 \\
NGC~6034 & B  & 3600 & 4 & 2.8 & 0.01 \\
NGC~6034 & g' & - & - & 1.4 & -0.03 \\
NGC~6034 & Rc & 1800 & 4 & 2.4 & 0.02 \\
NGC~6034 & r' & - & - & 1.4 & -0.19 \\
NGC~6034 & i' & - & & 1.1 & -0.01 \\
NGC~6034 & z' & - & - & 1.3 & -0.04 \\
\hline
1RXS J235814.4+150524 & u' & - & - & 1.1 & -0.02 \\
1RXS J235814.4+150524 & g' & - & - & 1.1 & 0.06 \\
1RXS J235814.4+150524 & Rc & 4500 & 1 & 2.8 & -0.11 \\
1RXS J235814.4+150524 & r' & - & - & 1.1 &  \\
1RXS J235814.4+150524 & i' & - & - & 1.2 & -0.22 \\
1RXS J235814.4+150524 & z' & - & - & 1.2 & -0.09 \\
\hline
RX J1119.7+2126 & u' & - & - & 1.4 & 0.08 \\
RX J1119.7+2126 & v & 49230 & 1 & 2.7 & 0.00  \\
RX J1119.7+2126 & B & 10800 & 1 & 3.0 & 0.00  \\
RX J1119.7+2126 & g & 18000 & 1 & 2.5 & -0.04   \\
RX J1119.7+2126 & g' & - & - & 1.8 & -0.15 \\
RX J1119.7+2126 & V & 7200 & 1 & 3.1 & 0.06  \\
RX J1119.7+2126 & Rc & 4500 & 1 & 2.1 & 0.29  \\
RX J1119.7+2126 & r & 10800 & 1 & 2.3 & 0.30  \\
RX J1119.7+2126 & r' & - & - & 1.7 & -0.08  \\
RX J1119.7+2126 & Ic & 7200 & 1 & 3.4 & -0.05  \\
RX J1119.7+2126 & i' & - & - & 1.4 & -0.18  \\
RX J1119.7+2126 & z' & - & - & 1.3 & -0.11 \\
\hline
\end{tabular}
\label{tab:imaging}
\end{table*}

These data map squares of $\sim 2.0\times 2.0$,
 $\sim 0.8\times 0.8$, and $\sim 0.8\times 0.8$ Mpc$^2$ for 
1RXS J235814.4+150524, RX J1119.7+2126, and NGC~6034, respectively.

We computed virial radii for the three groups from the Carlberg et 
al. (1997) formula, selecting as group members the galaxies in the
[0.059, 0.06134] redshift interval for RX J1119.7+2126, in the [0.17, 0.185] 
and [0.177, 0.181] redshift intervals for 1RXS J235814.4+150524, and in the 
[0.032, 0.037] redshift interval for NGC~6034.

- The virial radius for 1RXS J235814.4+150524 was then estimated to be 0.58 or 
1.31 Mpc (see below), so our images radius samples at least 76$\%$ of 
the virial area.

- The virial radius for RX J1119.7+2126 was estimated to be 0.72 Mpc,
so our images radius reaches about 0.56 times the virial radius.

- The virial radius for NGC~6034 was estimated to be 0.71 Mpc, so our images 
radius reach about 0.56 times the virial radius.

In any case, the spatial area of interest (half the virial radius,
following Jones et al. 2003) was covered for all three groups.

Data reduction was made with the MIDAS, Scamp, and Swarp (Bertin et
al. 2002, Bertin 2006) packages in the same way as in Guennou et
al. (2010). The resulting internal astrometric error is about
0.6~arcsec, well below the seeing of the considered images. Images in
different bands were aligned at the pixel scale and object catalogs
were extracted with SExtractor (Bertin $\&$ Arnouts 1996) in double
image mode.

\subsubsection{Star-galaxy separation}

To distinguish between stars and galaxies, we used the plot
of central surface brightness versus total magnitude shown in 
Fig.~\ref{fig:stargal}, which shows the data from the g' band SDSS images. 
This is an improvement on the results of Adami et al. (2007b) because 
OHP data were not very well adapted to perform star-galaxy separation
because of poor seeing.

We were then able to efficiently distinguish galaxies from stars down
to g'=20.4 (see Fig.~\ref{fig:stargal} for NGC~6034). We assumed that
all fainter objects were galaxies, according to the results of Adami et al. 
(2007b).

\begin{figure}
\caption[]{Central surface brightness versus total magnitude for the
  NGC~6034 field in the g' band. Red dots are assumed to be galaxies
  and black dots are stars.}
\label{fig:stargal}
\end{figure}

\subsubsection{Completeness of the catalogs}

\begin{figure}
\centering \mbox{\psfig{figure=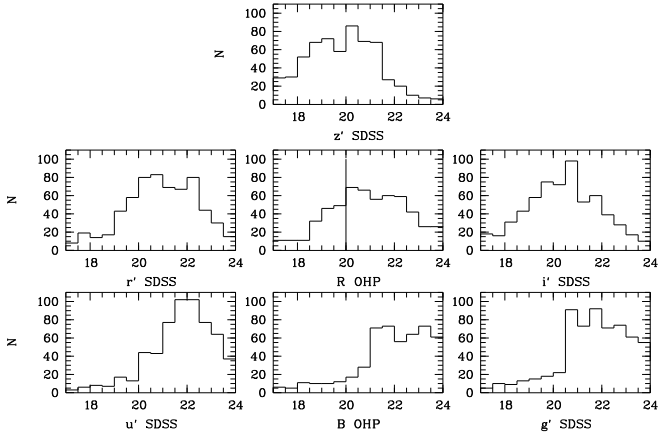,width=9cm,angle=270}}
\caption[]{Galaxy magnitude histograms from available images for
  NGC~6034. The adopted magnitude limit in the Rc band is shown.}
\label{fig:comp1}
\end{figure}

\begin{figure}
\centering \mbox{\psfig{figure=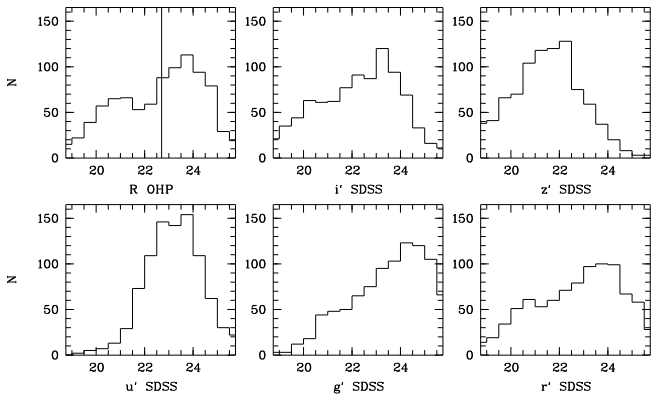,width=9cm,angle=270}}
\caption[]{Galaxy magnitude histograms from available images for 1RXS
  J235814.4+150524. The adopted magnitude limit in the Rc band is shown.}
\label{fig:comp2}
\end{figure}

\begin{figure}[h]
\centering \mbox{\psfig{figure=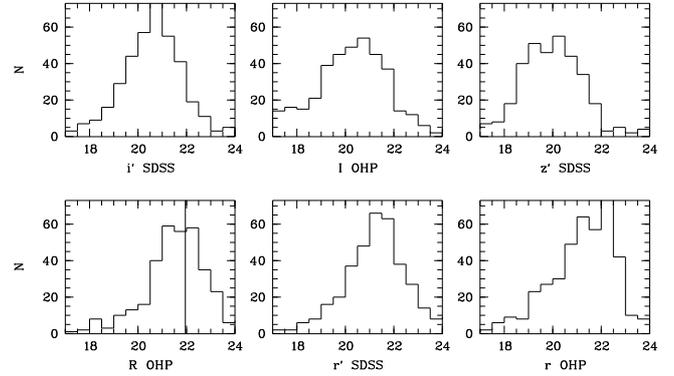,width=9cm,angle=270}}
\centering \mbox{\psfig{figure=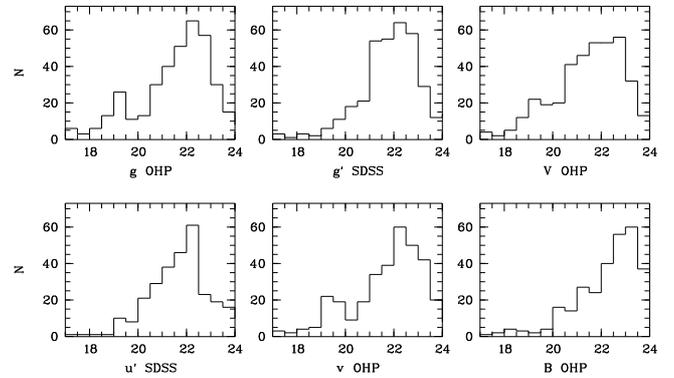,width=9cm,angle=270}}
\caption[]{Galaxy magnitude histograms from available images for RX
  J1119.7+2126. The adopted magnitude limit in the Rc band is shown.}
\label{fig:comp3}
\end{figure}

The completeness levels of the B and Rc imaging data were already given
in Adami et al. (2007b).  We show in Figs.~\ref{fig:comp1},
\ref{fig:comp2}, and \ref{fig:comp3} the magnitude histograms of all
available images. The adopted magnitude limits are
also shown in the Rc band. This limit was computed considering the 
position of the
peak of the magnitude histogram in each band, translating the value into
the Rc band using the mean color, and setting the magnitude limit to
that imposed by the shallowest band (most of the time u' SDSS).

\subsubsection{Photometric redshifts}

One of the
primary technical goals of the paper was to compute photometric
redshifts for galaxies along the lines of sight of the considered
groups of galaxies.  This technique of defining a structure population
with photometric redshifts is becoming common (e.g. Adami et al. 2008) 
but this is one of the first times to our knowledge that it is applied to 
fossil groups.

We could have directly used SDSS public photometric redshifts.
However, these photometric redshifts are only based on a five-band
survey and the number of available spectroscopic redshifts to check
and adjust the photometric redshift values is quite limited in this
survey for the presently considered areas, especially toward faint 
magnitudes. This is why we 
collected data both in additional photometric bands and deep
spectroscopy. This gave us a six band coverage for 1RXS
J235814.4+150524, seven band for NGC~6034, and twelve band for RX
J1119.7+2126. We note that the OHP R band data for 1RXS
J235814.4+150524 were observed under variable sky conditions and 
calibration remained uncertain. We therefore chose to not use them
when computing photometric redshifts.
We then used the Le~Phare code (e.g. Ilbert et
al. 2006). Briefly, this photometric redshift code is able to compare
observed magnitudes with predicted ones created by templates from the
literature. We then selected the CE templates (see Arnouts et
al. 1999) when computing simple photometric redshifts. In addition to
the photometric redshifts, this gave us an estimate of the galaxy
photometric type through an integer number between 1 and 66. In the
present paper, the interval $1-20$ was assigned to early-type
galaxies (elliptical galaxy templates), $20-35$ to early-spiral 
galaxies (Sbc galaxy templates), and $35-66$ to late-spiral 
galaxies (Scd and Irr galaxy templates). We also selected the
Bruzual $\&$ Charlot (2003, BC03 hereafter) templates to compute
stellar population ages for galaxies with available spectroscopic
redshifts. Finally, we adopted the Calzetti $\&$ Heckman (1999)
extinction law.

Le~Phare was also able to estimate possible shifts in photometric band
zero points, by comparing the photometric and spectroscopic redshifts
used for training sets. Shifts were computed fixing photometric
redshifts to the spectroscopic values and averaging the residuals in
each of the bands. This technique is crucial when using photometry
from observatories where the sky conditions are not always very
good. We list these shifts in Table~\ref{tab:imaging}. They allow us
to take into account internal photometry inhomogeneities between
different bands.

\begin{figure}
\centering \mbox{\psfig{figure=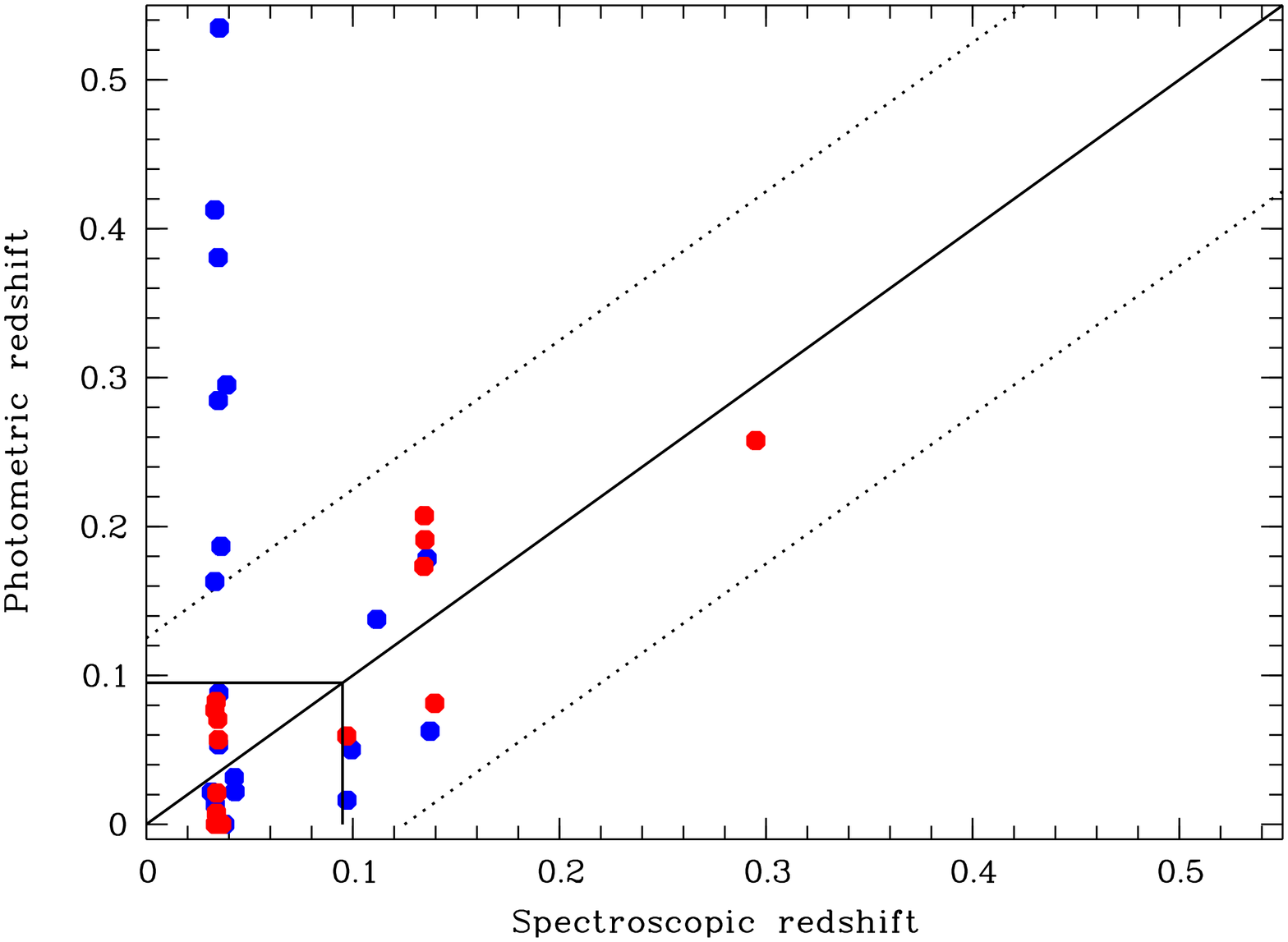,width=8cm,angle=270}}
\centering \mbox{\psfig{figure=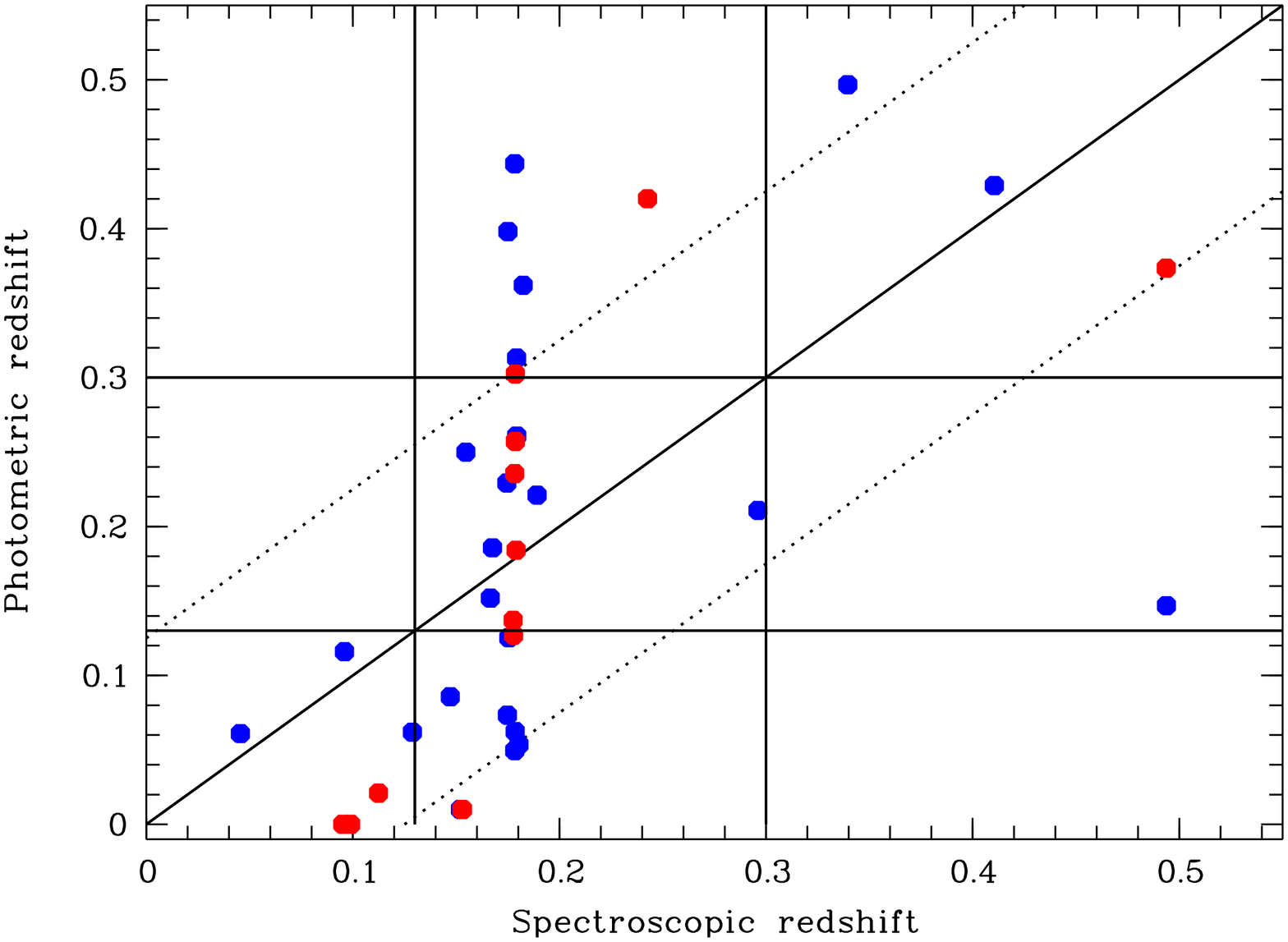,width=8cm,angle=270}}
\centering \mbox{\psfig{figure=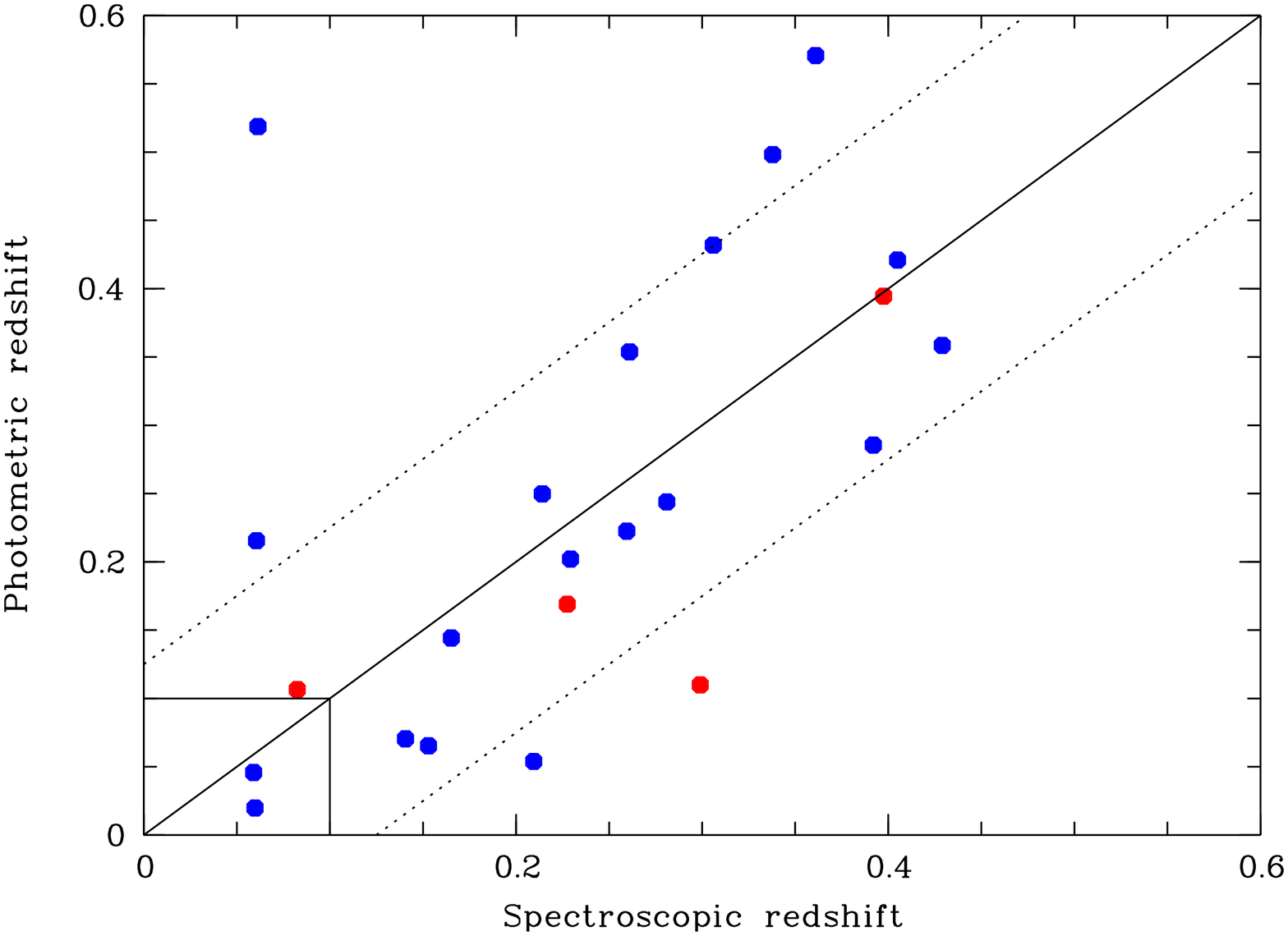,width=8cm,angle=270}}
\caption[]{Spectroscopic versus photometric redshifts for the NGC~6034
  group (top), the 1RXS J235814.4+150524 fossil group (middle), and
  the RX J1119.7+2126 fossil group (bottom). Red dots are early-type
  galaxies, while blue dots are spiral galaxies (see text). The
  inclined lines show the perfect relation and its $\pm$0.125
  1$\sigma$ uncertainty.  We also show the region chosen as the
``structure'' redshift range.}
\label{fig:zphot}
\end{figure}

Fig.~\ref{fig:zphot} shows the resulting comparisons between
photometric and spectroscopic redshifts for the three
groups. Given the spectroscopic catalogs we collected for
our clusters (see Fig.~\ref{fig:CMR} with measured galaxies down to
Rc$\sim$17.5 for NGC~6034 and down to Rc$\sim$21 for the two fossil 
groups), Fig.~\ref{fig:zphot} gives a
relatively good knowledge of the quality of our photometric redshifts
over nearly the entire useful magnitude range.  The typical
dispersions in the photometric/spectroscopic redshift relations are
0.13 for 1RXS J235814.4+150524, 0.1 for RX J1119.7+2126, and 0.06 for the
NGC~6034 group when excluding the cluster redshift interval. We define in
Fig.~\ref{fig:zphot} the optimum photometric redshift intervals
excluding as many field (fore- or back-ground) galaxies 
as possible and keeping as many
structure galaxies as possible.  The intervals are [0.13, 0.3] for
1RXS J235814.4+150524, [0., 0.095] for the NGC~6034 group, and [0.,
0.1] for RX J1119.7+2126. These intervals are shown in 
Fig.~\ref{fig:zphot} on the vertical (photometric redshifts) and
horizontal (spectroscopic redshifts) axes. On the vertical axes, they show the 
selection we applied to the photometric redshift catalogs. On the 
horizontal axes they illustrate that we still consider a few galaxies as group
members that are fore- or back-ground objects on the basis of
their spectroscopic redshift.

The spectroscopic catalogs allow us to estimate 
that we lose $\sim$40$\%$ of structure galaxies in this way and that we
only include $\sim$10$\%$ of field galaxies in 1RXS J235814.4+150524. The
percentage of lost structure members is important, but $\sim$80$\%$ of
these lost galaxies are late-type objects. This means that we are able
to distinguish efficiently between early-type galaxies that lie inside
and outside of the [0.13, 0.3] redshift interval for 1RXS
J235814.4+150524. This tendency to have degraded photometric redshift
estimates for relatively bright cluster late-type galaxies was already
shown in Guennou et al. (2010) and is probably caused by a lack of literature 
late-type templates adapted to high-density environments.

For the NGC~6034 group, we also lose $\sim$30$\%$ of structure
galaxies and include $\sim$10$\%$ of field galaxies. Similarly to 1RXS
J235814.4+150524, all lost galaxies are late-type objects, so we are
able to efficiently distinguish between early-type galaxies that lie
inside and outside of the [0., 0.095] redshift interval for the
NGC~6034 group.

It is not possible to reliably estimate the percentages of lost
structure galaxies and included field galaxies for RX J1119.7+2126,
because of the few spectroscopic redshifts lying inside the
structure, so we chose the same limitations as for NGC~6034. This
proved to give consistent results when computing luminosity functions
estimated with photometric redshifts and with statistical subtraction
techniques.

\section{Nature of the studied groups}

The first question is to know if we are indeed dealing with fossil
groups in our sample, since partial data can sometimes lead to wrong
interpretations of the nature of a given group. For example, NGC~6034
was claimed to be a fossil group by Yoshioka et al. (2004). However,
considering new optical data, Lopes de Oliveira et al. (2010) showed
that this group was not a member of the fossil class. 

We first confirm that the NGC~6034 galaxy is not associated with a
fossil group. From available spectroscopic redshifts in the
literature, two group member galaxies are less than 2 magnitudes
fainter than NGC~6034 and are located at 185 and 352 kpc from this
galaxy. Lopes de Oliveira et al. (2010) estimated the virial radius of
this group to be 940 kpc. The two mentioned galaxies are therefore inside
half the virial radius.

RX J1119.7+2126 was classified as a fossil group by Jones et
al. (2003).  The spectroscopic data of the present paper, already partly used in
Adami et al. (2007b), show that besides the dominant galaxy, the
second brightest galaxy member of the group and within a
circle of radius 360~kpc is 2.53 magnitudes fainter in the R
band. There is only one galaxy less than 2 magnitudes fainter than the
dominant galaxy, without spectroscopic redshift, which is potentially a
group member from photometric redshift results. This galaxy is 1.94
magnitude fainter than the dominant galaxy and is outside half the
virial radius of the group.  Finally, the X-ray luminosity is high
enough to classify RX J1119.7+2126 as a fossil group following
Jones et al. (2003).

1RXS J235814.4+150524 is luminous enough in X-rays (Santos et
al. 2007) to be potentially considered as a fossil group. The
velocity dispersion of this group is at most 578 km/s considering
our spectroscopically measured galaxies between z=0.17 and 0.185 (or 254 km/s
when considering galaxies between z=0.177 abd z=0.181). With 
the virial radius definition
of Carlberg et al. (1997), this corresponds to 1.31 Mpc.  With the
spectroscopic catalogs in hand, there are only two group member
galaxies in the available field of view that are less than two
magnitudes fainter than the dominant galaxy. The second-brightest
galaxy in our field of view is at 1.015 Mpc from the dominant galaxy;
it is therefore beyond half the virial radius and cannot be used to
invalidate the fossil nature of the considered group following Jones 
et al. (2003). The third-brightest 
galaxy of RXJ2359 is very close to the central galaxy and
the magnitude difference between these two galaxies is 1.8 magnitudes
in the Rc band.  

From the photometric redshift catalog, there are
three potential group member galaxies in the available field of view
which are less than two magnitudes fainter than the dominant
galaxy. The second brightest of these galaxies is not a member galaxy (this 
galaxy has a spectroscopic redshift outside of the group). The two next 
galaxies are at more than 750 kpc from the dominant galaxy and are 
therefore beyond half the virial radius.

At the end, only the third brightest galaxy in the spectroscopic
sample is bright and close enough to the dominant galaxy to violate the
fossil group definition. However, given the difference of nearly two magnitudes,
we can say that RXJ2359 is very close to the fossil group status from
the Jones et al. (2003) definition (see also Section 5.5 of the recent
paper by Proctor et al. 2011).

\section{Large-scale environment of the fossil groups}
 
On the one hand, we have already shown in Adami et al. (2007b) that RX
J1119.7+2126 is very isolated from the cosmic web. On the other hand,
Lopes de Oliveira et al. (2010) showed that NGC~6034 is highly
connected to the cosmic web with at least twelve clusters of galaxies
at similar redshifts in a two-degree field.

\begin{figure}
\centering \mbox{\psfig{figure=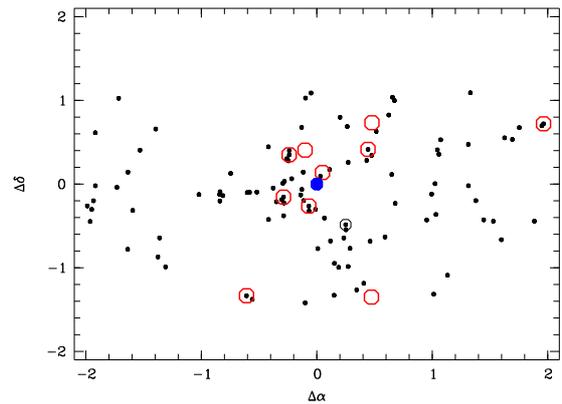,width=8cm,angle=270}}
\caption[]{Map of the large-scale environment of the 1RXS
  J235814.4+150524 fossil group (blue disk). Black dots are the known
  galaxies within the redshift range [0.16;0.195]. The black circle
  shows the only galaxy less than two magnitudes fainter than the
  dominant galaxy. Red circles are the known compact galaxy structures
  (groups or clusters) in the same redshift range. Units of the figure
  are decimal degrees relative to the group center. }
\label{fig:LSS2359}
\end{figure}

Fig.~\ref{fig:LSS2359} shows the large-scale environment of the 1RXS
J235814.4+150524 fossil group in a $\sim$50 Mpc area. We first note
that ten compact structures of galaxies (groups or clusters from NED)
are known in this area in the redshift range [0.16;0.195].  1RXS
J235814.4+150524 is therefore much less isolated than RX
J1119.7+2126. However, only one galaxy in this area is less than two
magnitudes fainter than the dominant galaxy of 1RXS J235814.4+150524 (so 
even fainter than the considered condition to be only fainter than the 
dominant galaxy). As demonstrated in Adami et al. (2007b), this
probably means that 1RXS J235814.4+150524 is the dominant structure in
this region and that the ten compact galaxy structures detected are
probably very low mass systems.  1RXS J235814.4+150524 is therefore
not acting as a negligible system in its cosmological bubble, as
was the case for the NGC~6034 group, which is minor compared to the giant 
clusters in its vicinity.

We conclude that the fossil group 1RXS J235814.4+150524 is only
embedded in a relatively diffuse environment, therefore not providing
many infalling galaxies.

\section{Small-scale properties of the groups}

We will now investigate the properties of the faint galaxy populations of the 
present group sample.

\subsection{Spatial distributions of the group members}

Lopes de Oliveira et al. (2010) showed density contours of the likely
members of the NGC~6034 group in their figure 3. However, this plot is
probably partially polluted by field galaxies. Here, we selected
likely group members by considering only galaxies with photometric
redshifts inside the intervals defined earlier. As stated before, this
will mainly give the distribution of the early-type galaxies of the
group, since we lose a considerable part of the late-type galaxies. We also
note that several field galaxies will still be included in the group
population because photometric redshifts are not as precise as
spectroscopic ones. However, following Ilbert et al. (2005), we
estimated that this contribution is small for the three considered
groups (less than one galaxy per half magnitude and per deg$^2$). We
then computed a galaxy density map from this sample with an adaptive
kernel code as described in Adami et al. (2007a).  The result is shown
in Fig.~\ref{fig:SS6034}. We clearly see the main part of the
structure as well as a north-east extension correlated with a
relatively bright galaxy (a group member) and in the direction of the
Abell~2151 complex. These results are consistent with the picture
drawn by Lopes de Oliveira et al. (2010) in which the NGC~6034 group is
part of a larger complex that includes several clusters of galaxies. 

\begin{figure}
\caption[]{Adaptative kernel galaxy density map of galaxies with
  photometric redshifts lower than 0.095 along the NGC~6034 line of
  sight (blue contours). The first contour corresponds to the mean 
galaxy density value over the field of view, and the last contour
to the maximal value. The size of the image is
  20$\times$20~arcmin$^2$. North is up and east is to the left. 
  Red circles are galaxies with a
  spectroscopic redshift inside the group (small and large circles are
  galaxies with mean stellar population younger and older than
  10$^{10}$ yr). White squares are galaxies with spectroscopic
  redshift outside the group. Green lines represent the directions to
  the closest galaxy clusters (see Lopes de Oliveira et al. 2010).}
\label{fig:SS6034}
\end{figure}

Fig.~\ref{fig:SS2359} shows the galaxy density map for galaxies with
photometric redshifts in the z=[0.13, 0.3] range along the 1RXS
J235814.4+150524 line of sight. Once again, we clearly detect the core
of the fossil group. Density contours are less extended than for the
NGC~6034 group but seem to show a south extension nicely correlated with
the direction of three close compact galaxy structures. 
We therefore at least partially confirm the
results of the previous section: 1RXS J235814.4+150524 appears to be a
relatively isolated structure, but with a clear signature of the
surrounding cosmic web.

\begin{figure}
\caption[]{Adaptative kernel galaxy density map of galaxies with
  photometric redshifts in the [0.13, 0.3] redshift interval along the
  1RXS J235814.4+150524 line of sight (blue contours). The first contour 
corresponds to the mean galaxy density value over the field of view, and the 
last contour to the maximal value. The size of the
  image is 11.6$\times$11.6~arcmin$^2$. North is up and east is to the 
  left. Symbols are the same as in
  Fig.~\ref{fig:SS6034}.  Green lines represent the direction of the
  closest known compact structures of galaxies from NED.}
\label{fig:SS2359}
\end{figure}

We already computed a member galaxy density map for RX J1119.7+2126 in
Adami et al. (2007b) based on a (B-Rc) color selection. We present
here the new results based on the 12-band photometric redshift
analysis (Fig.~\ref{fig:SS1119}). The density map computed here
is very similar to Fig. 6 of Adami et al. (2007b), also showing the
north and east concentrations, which confirms that the immediate vicinity
of the dominant galaxy is strongly depopulated in terms of galaxies.

\begin{figure}
\caption[]{Adaptative kernel galaxy density map of galaxies with
  photometric redshifts in the [0.,0.1] redshift interval along the
  RX J1119.7+2126 line of sight (blue contours). The first contour 
corresponds to the mean galaxy density value over the field of view, and the 
last contour to the maximal value. The size of the image
  is 11.5$\times$11.5~arcmin$^2$. North is up and east is to the left.  
  Symbols are the same as in
  Fig.~\ref{fig:SS6034}. The fossil group main galaxy is at the image center.}
\label{fig:SS1119}
\end{figure}

To confirm the isolation status of the fossil groups compared
to the NGC~6034 group, we also searched for recent infalling activity
through substructure analysis. This is not possible for RX
J1119.7+2126 because we have detected only five spectroscopic
members. For 1RXS J235814.4+150524, which is not very massive 
and is the dominant structure of its area (see previous
section), we expect to find only minor substructuring, indicative of a
weak infalling activity. We performed a Serna-Gerbal substructure
search (Serna $\&$ Gerbal 1996) on the spectroscopic catalog of 1RXS
J235814.4+150524 and detected no substructure.  With the available
spectroscopic catalog, major substructures should be detected if
present (see e.g. Adami et al. 2011). As a comparison, we detected a
potential substructure in the NGC~6034 group with an estimated virial mass of 4
10$^{12}$ M$_\odot$, typically one tenth of the mass of the parent
group itself (see Lopes de Oliveira et al. 2010). The main galaxy
of this substructure is the central galaxy of the secondary peak 
located northeast of Figs.~\ref{fig:SS6034}. The NGC~6034 group
therefore seems to have accreted a very small group coming from the
surrounding cosmic web, while 1RXS J235814.4+150524 does not show a
similar behavior.

\begin{figure}
\centering \mbox{\psfig{figure=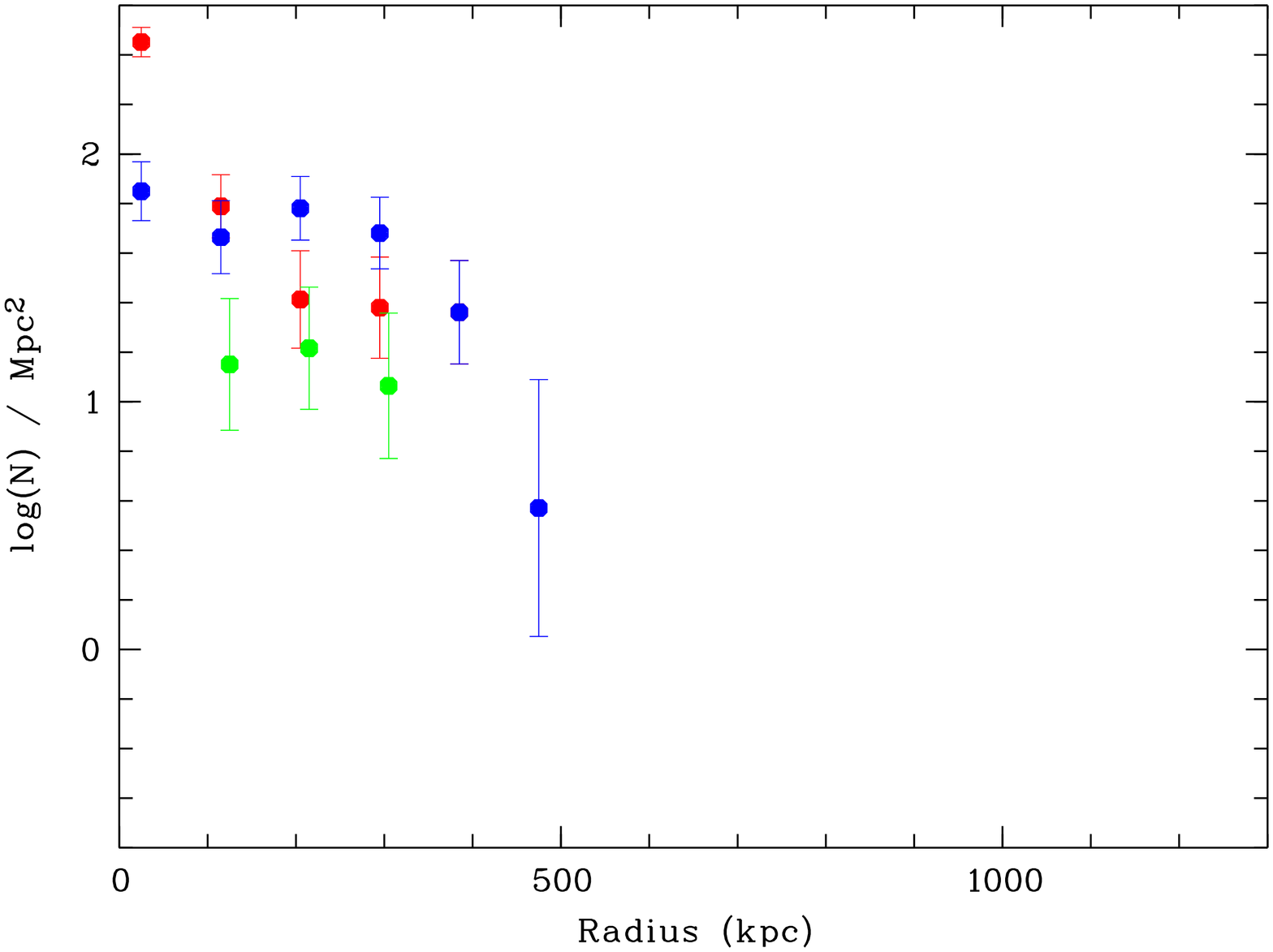,width=8cm,angle=270}}
\centering \mbox{\psfig{figure=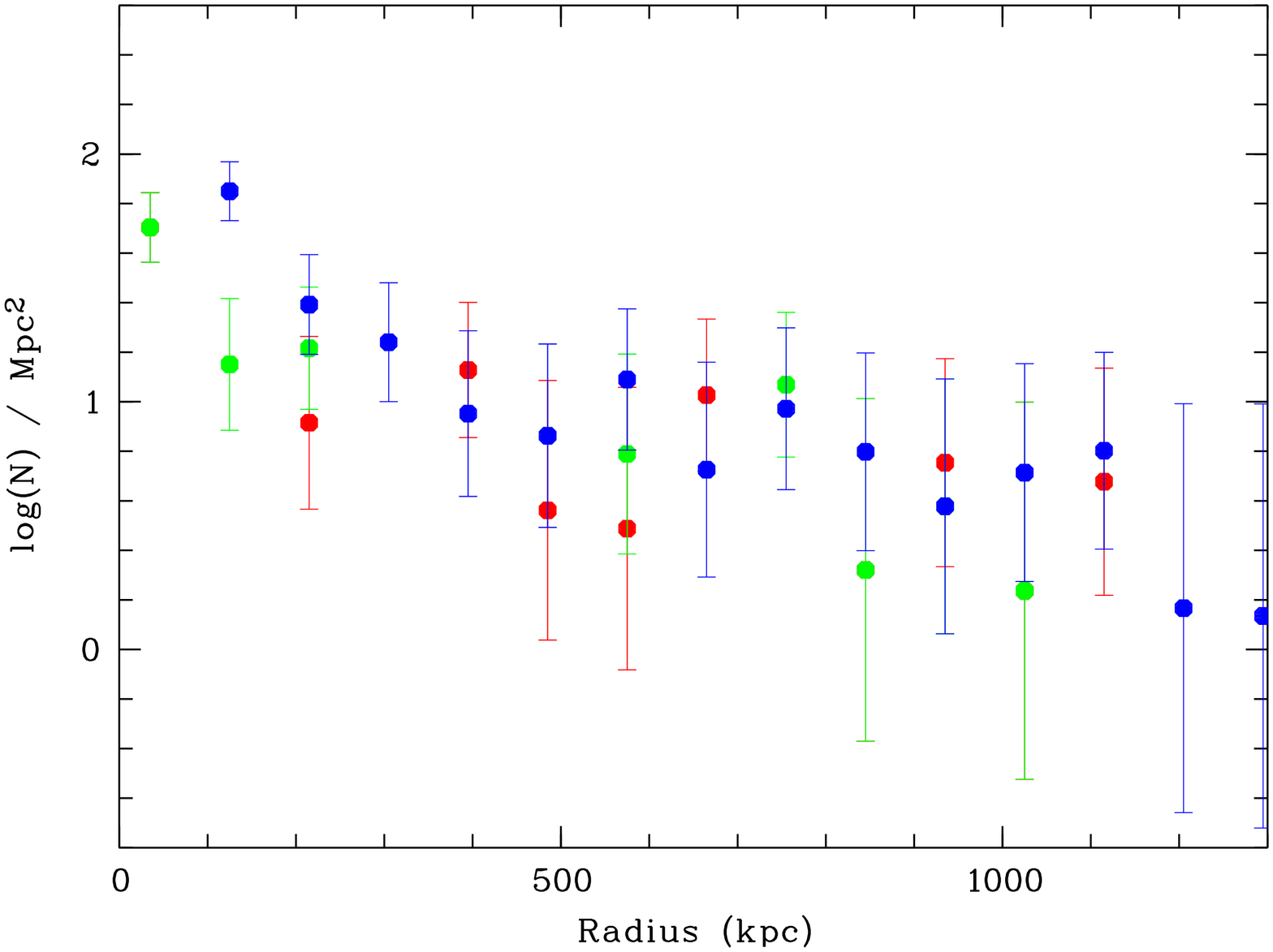,width=8cm,angle=270}}
\caption[]{Galaxy density profiles of elliptical (red dots), early
  spiral (green dots), and late spiral (blue dots) galaxies for 
  NGC~6034 (top) and 1RXS J235814.4+150524 (bottom). Bin size for the
two figures was 90 kpc.}
\label{fig:densprof}
\end{figure}

The photometric redshift computation process also produces estimates
of the photometric type of the galaxies, as already stated. Keeping in
mind that we are dealing with catalogs of potential structure members 
biased toward early-type galaxies, we computed galaxy density profiles
for the considered groups as a function of  galaxy type
(Fig.~\ref{fig:densprof}).  Since galaxy density contours in
Figs.~\ref{fig:SS6034} and ~\ref{fig:SS2359} are relatively concentrated and
isotropic, we computed numbers of galaxies within annuli for these two
groups, limiting the galaxy samples to a common absolute magnitude
value given by the shallowest data. Only galaxies brighter than
M$_{Rc}=-16.85$ were kept. 

Fig.~\ref{fig:densprof} shows that the NGC~6034 and 1RXS
J235814.4+150524 groups show similar galaxy density profiles
with similar density values. There is one main difference, however:
while 1RXSJ235814.4+150524 basically shows a flat profile close
to the center, NGC~6034 is more cuspy, mainly because of elliptical-like 
galaxies.

These possible tendencies
are not taking into account magnitude effects, however. We
therefore compute the color-magnitude relations and the 
luminosity functions of the three groups.

\subsection{Color-magnitude relations}

\begin{figure}
\centering \mbox{\psfig{figure=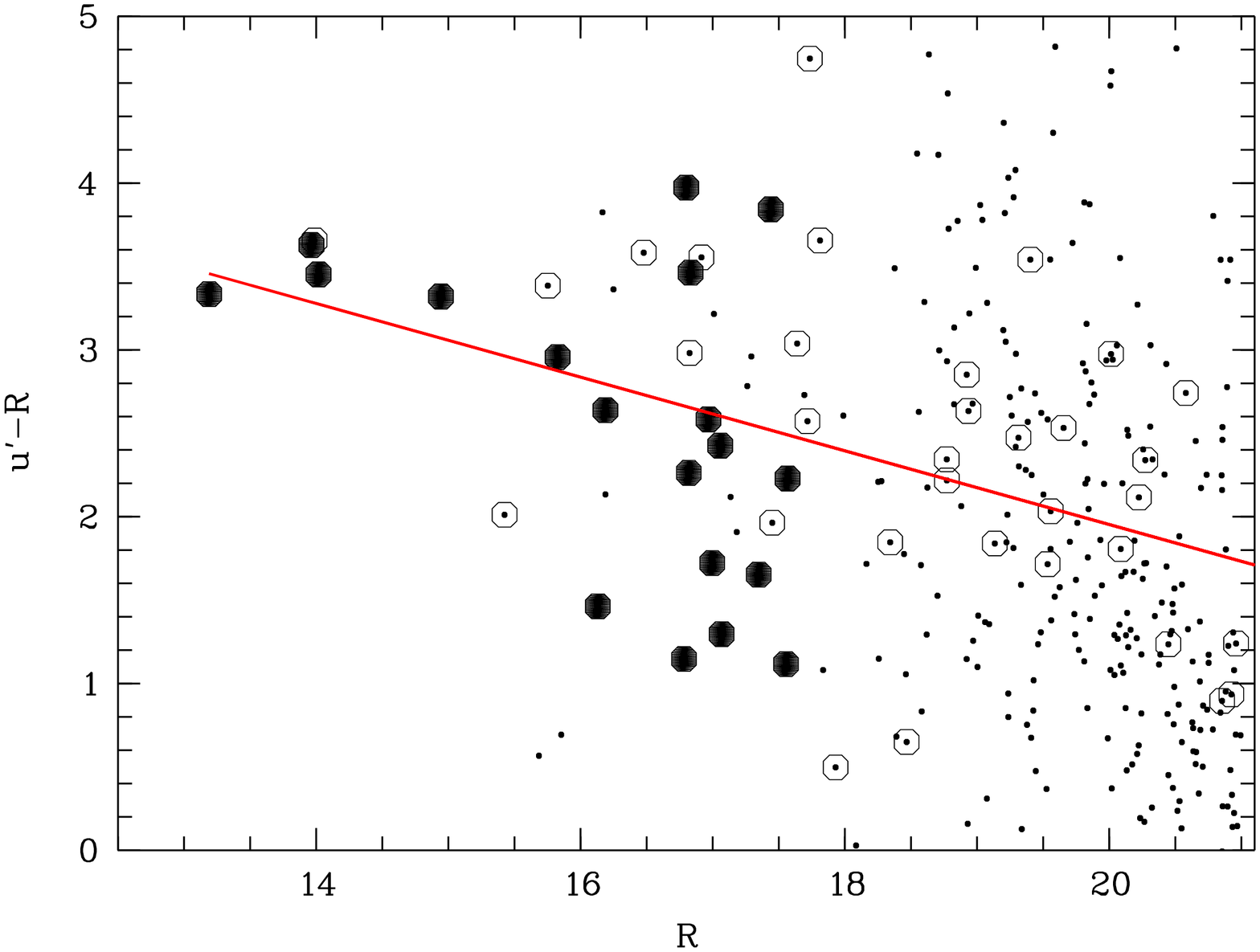,width=8cm,angle=270}}
\centering \mbox{\psfig{figure=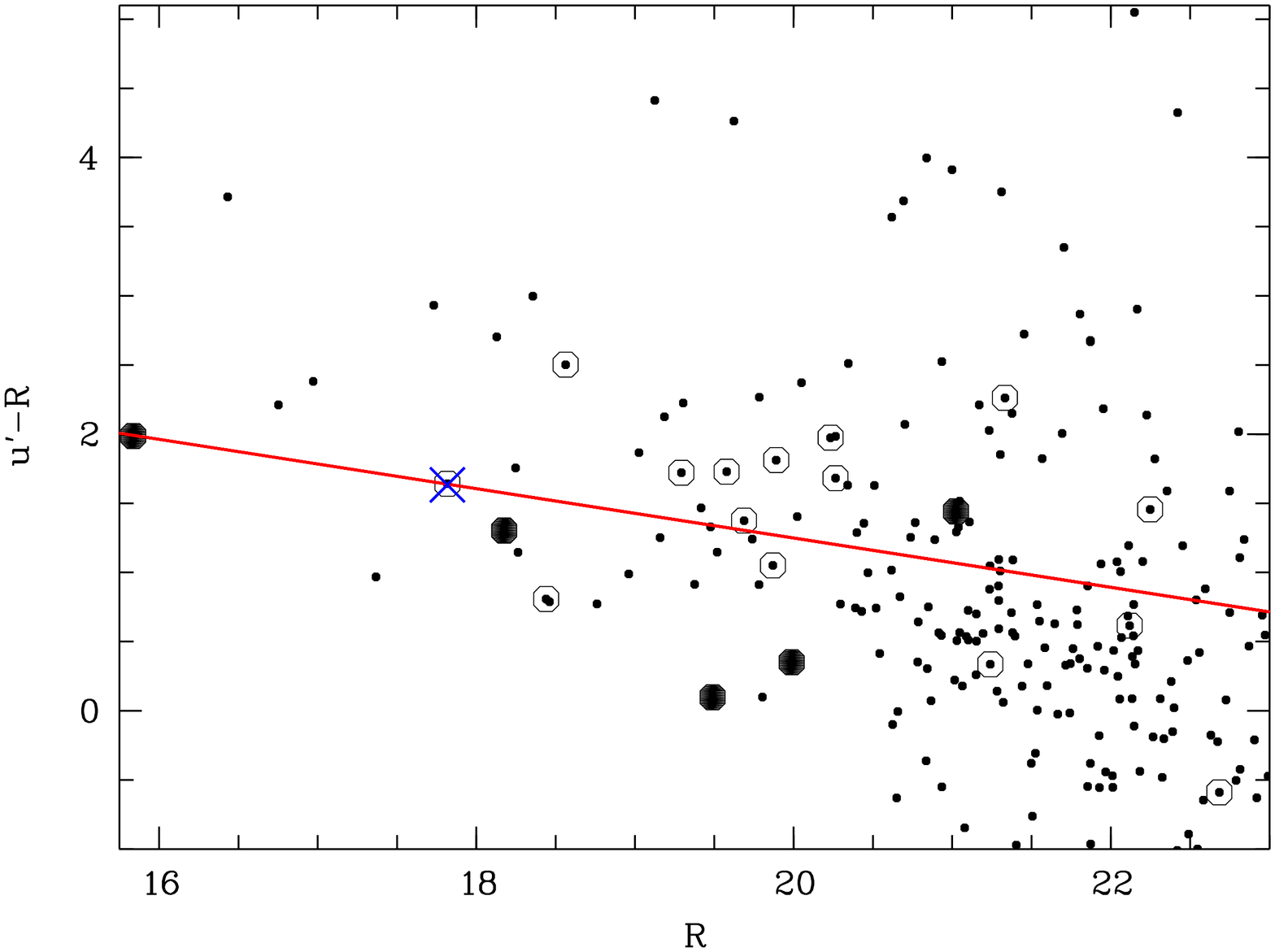,width=8cm,angle=270}}
\caption[]{Color-magnitude relations for NGC~6034 (top), 1RXS
  J235814.4+150524 (middle), and RX J1119.7+2126 (bottom). Dots: all
  galaxies detected along the lines of sight. Filled circles: galaxies
  spectrocopically classified as group members. Open circles: galaxies
  that potentially are members of the group from their photometric redshift.
  The blue crosses
  (middle and lower figures) are potential group  members (from
  photometric or spectroscopic redshifts) less than two magnitudes
  fainter than the dominant group galaxy but which are more distant
  than half the virial radius from the group center. They are
  therefore no indrance to classify the group as a fossil
  structure. The red lines are the mean red sequences computed
selecting spectroscopic group members and potential members through 
photometric redshifts.}
\label{fig:CMR}
\end{figure}

Color-magnitude relations (CMR hereafter) are very useful to sample
galaxy populations, in particular since cluster early-type galaxies 
most of the time exhibit a red sequence (RS hereafter) that is 
characteristic of their
evolution (e.g. Kodama $\&$ Arimoto 1997). We already showed such RS in 
the CMR of RX J1119.7+2126 in Adami et al. (2007b), based on
statistical arguments, and a similar RS was also detected in the
NGC~6034 group by Lopes de Oliveira et al.  (2010). We show in
Fig.~\ref{fig:CMR} the color-magnitude relations of the three
groups computed with the present data. RS are discernable in
the three cases, confirming that we are probably dealing with 
sufficiently old
structures that had time to assemble an old and early-type galaxy
population.

\subsection{Galaxy luminosity functions of the fossil groups}

We know that fossil groups only have a faint galaxy population (at least 
two magnitudes fainter than the brightest object) in addition to their 
brightest galaxy. The previous section considered galaxies in the
whole observed area. To check whether the considered fossil
groups in the present paper include galaxies in the dwarf regime or
not, we now compute luminosity functions in half the virial
radius. This is the area where faint galaxy populations could be the 
most affected.

\begin{figure}
\centering \mbox{\psfig{figure=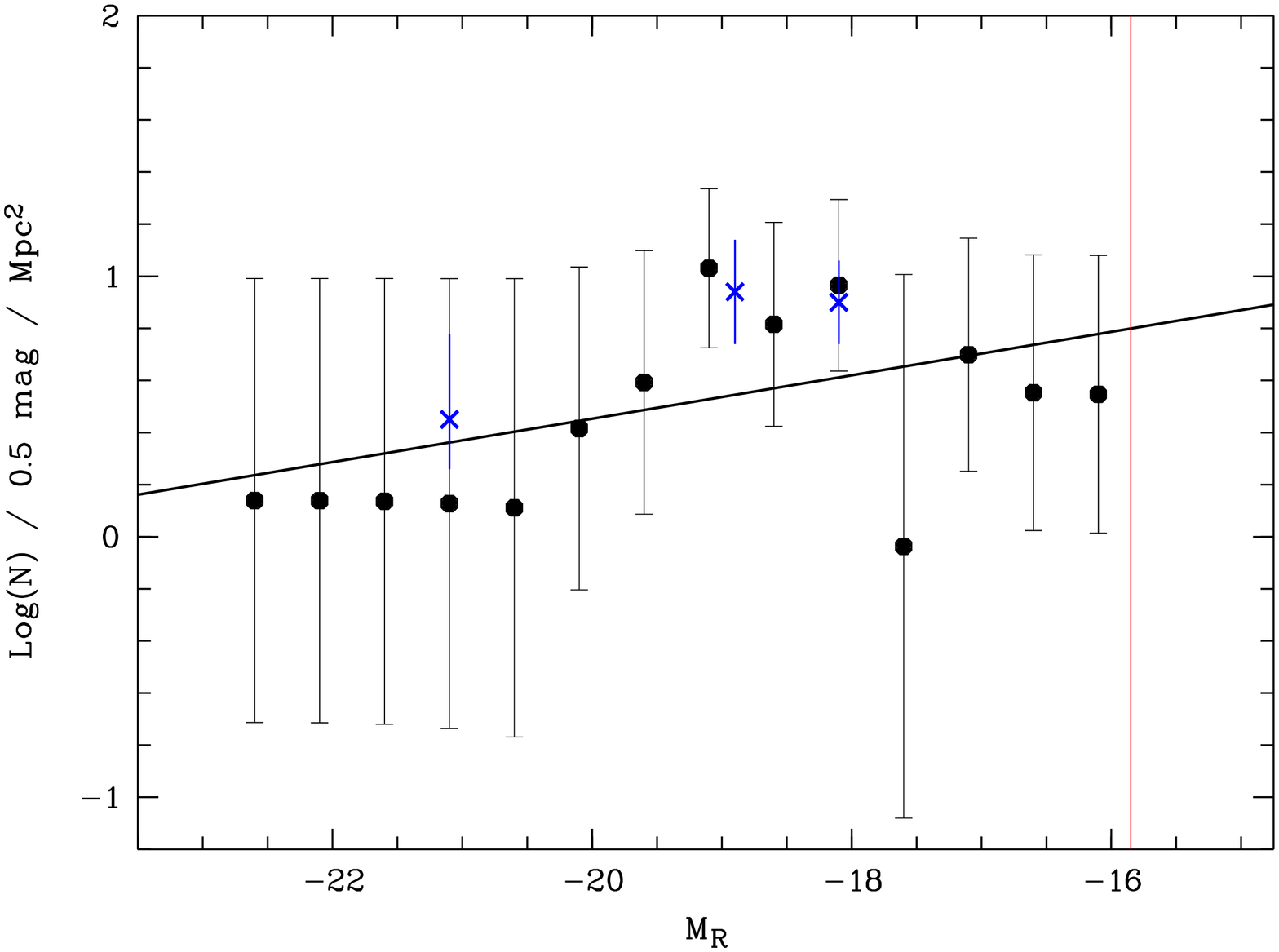,width=8cm,angle=270}}
\centering \mbox{\psfig{figure=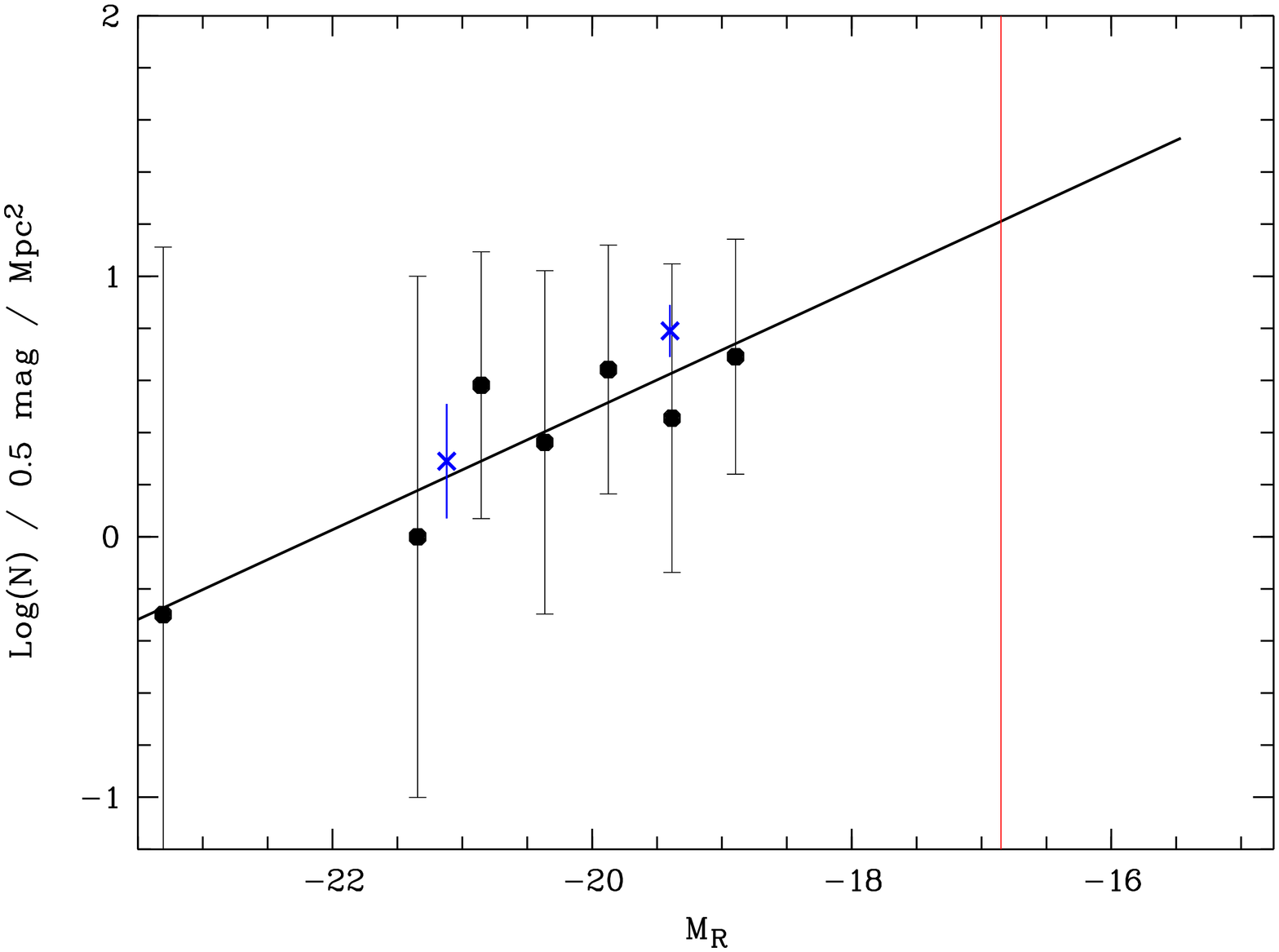,width=8cm,angle=270}}
\centering \mbox{\psfig{figure=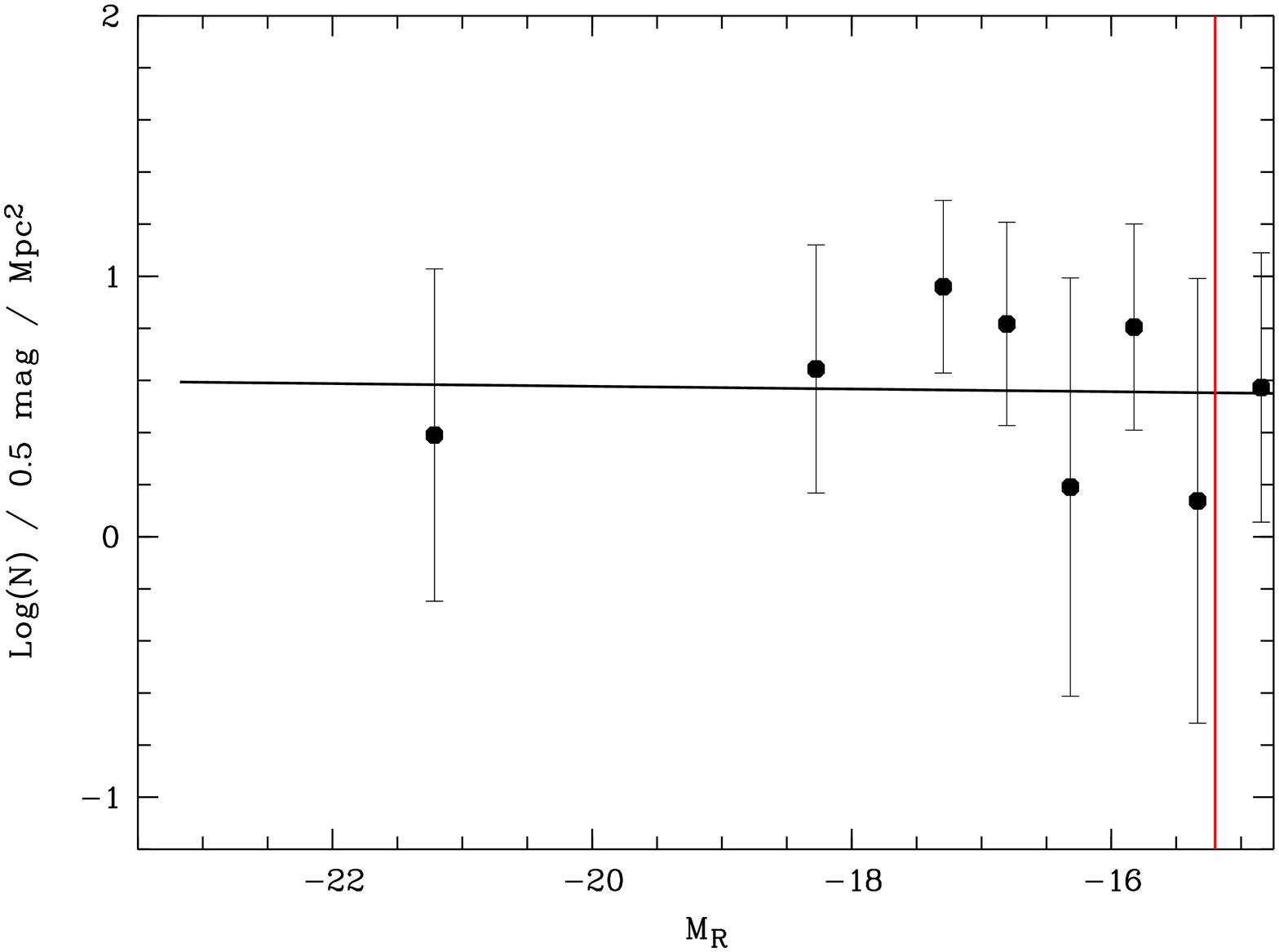,width=8cm,angle=270}}
\caption[]{Photometric redshift based Rc band galaxy luminosity
  functions of NGC~6034 (top), 1RXS J235814.4+150524 (middle), and RX
  J1119.7+2126 (bottom). The black lines are the mean luminosity
  functions of the M$_{Rc}$ intervals [-22.5, -16.]  (NGC~6034),
  [-22.5, -19.] (1RXS J235814.4+150524), and [-21.5, -15.25]  
(RX J1119.7+2126]. The red lines are the adopted
  completeness levels. The blue crosses are from the spectroscopic
  catalogs.}
\label{fig:FDLzphot}
\end{figure}

To compute luminosity functions in the Rc band for the three
groups, we first considered the photometric redshifts to
define the group membership, as we did for Coma (Adami et
al. 2008). Briefly, the photometric redshifts were used to remove as
many field galaxies as possible (with the limitations quoted in the
previous section). This was mainly efficient for early-type
galaxies. Then, we statistically subtracted the field luminosity
function in the same band and in the considered volume with the Ilbert
et al. (2005) estimates. We show the results in
Fig.~\ref{fig:FDLzphot}. On the one hand, we confirm the lack of faint
galaxies in 1RXS J235814.4+150524; we may have reached the luminosity
function turnover of this structure. On the other hand, the luminosity
functions of the NGC~6034 and RX J1119.7+2126 groups have more regular
shapes down to the completeness level, with a relatively flat slope, if
we except the expected dip between the dominant and the second-brightest
galaxy for RX J1119.7+2126.

\begin{figure}
\centering \mbox{\psfig{figure=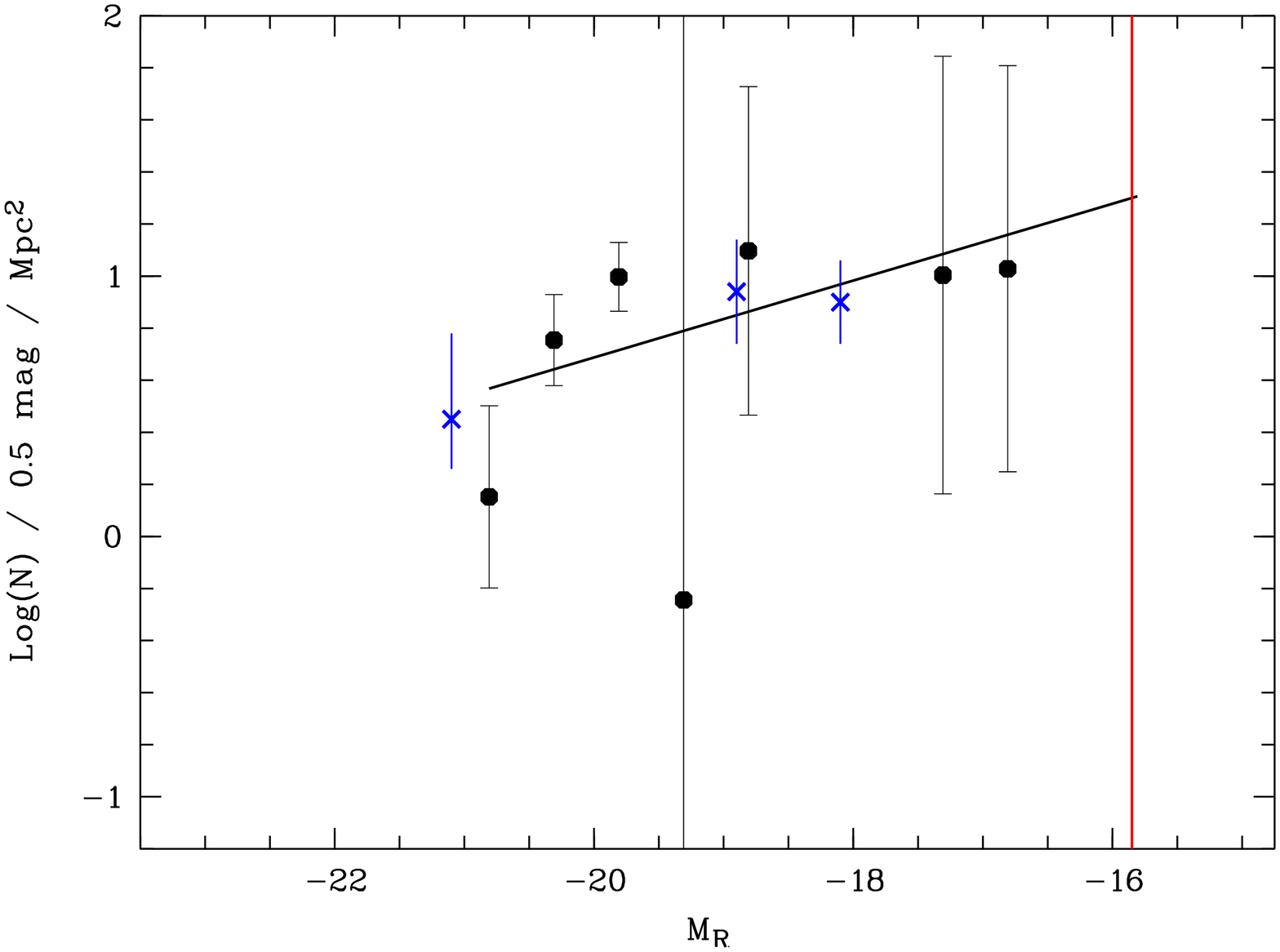,width=8cm,angle=270}}
\centering \mbox{\psfig{figure=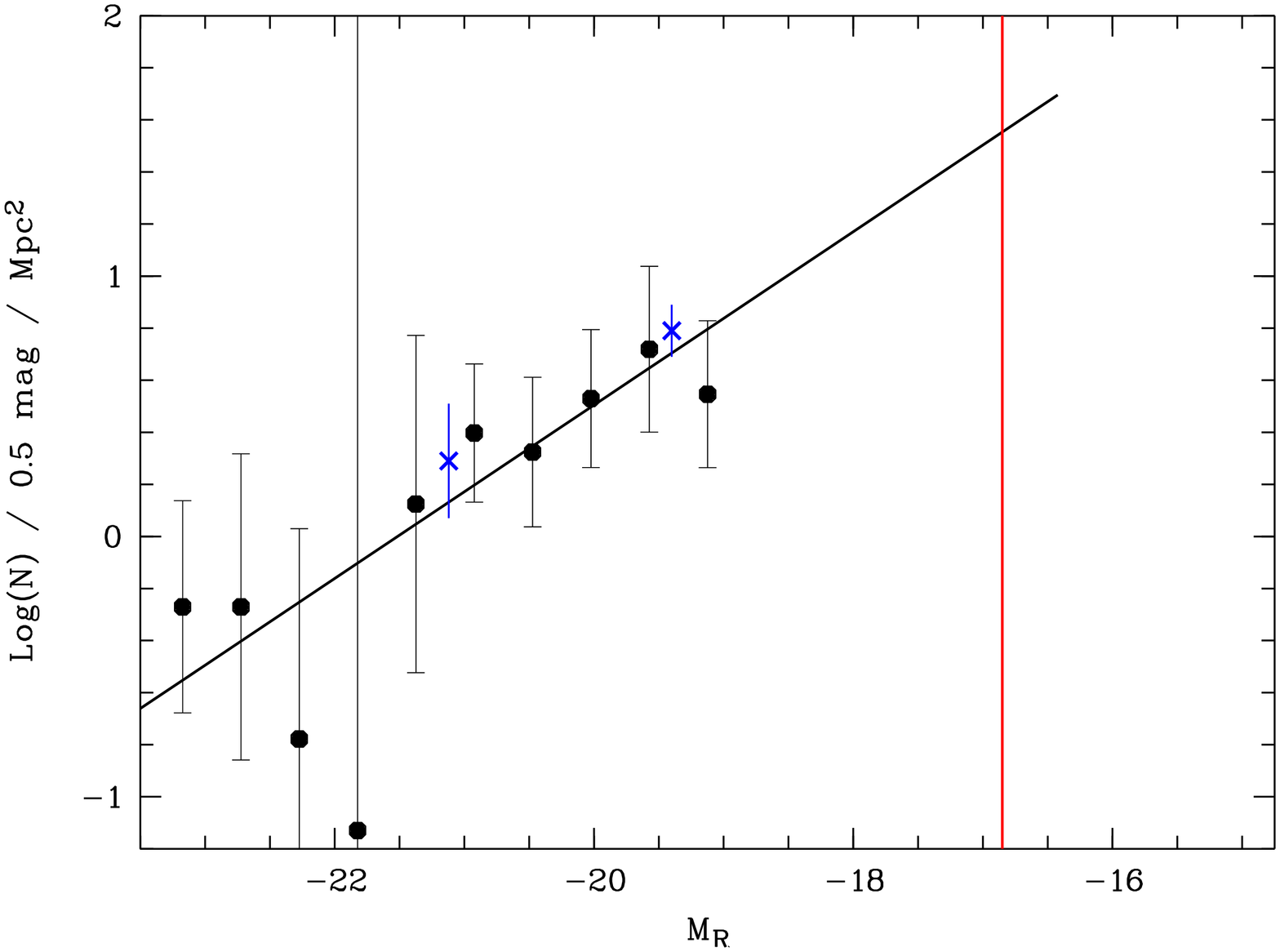,width=8cm,angle=270}}
\centering \mbox{\psfig{figure=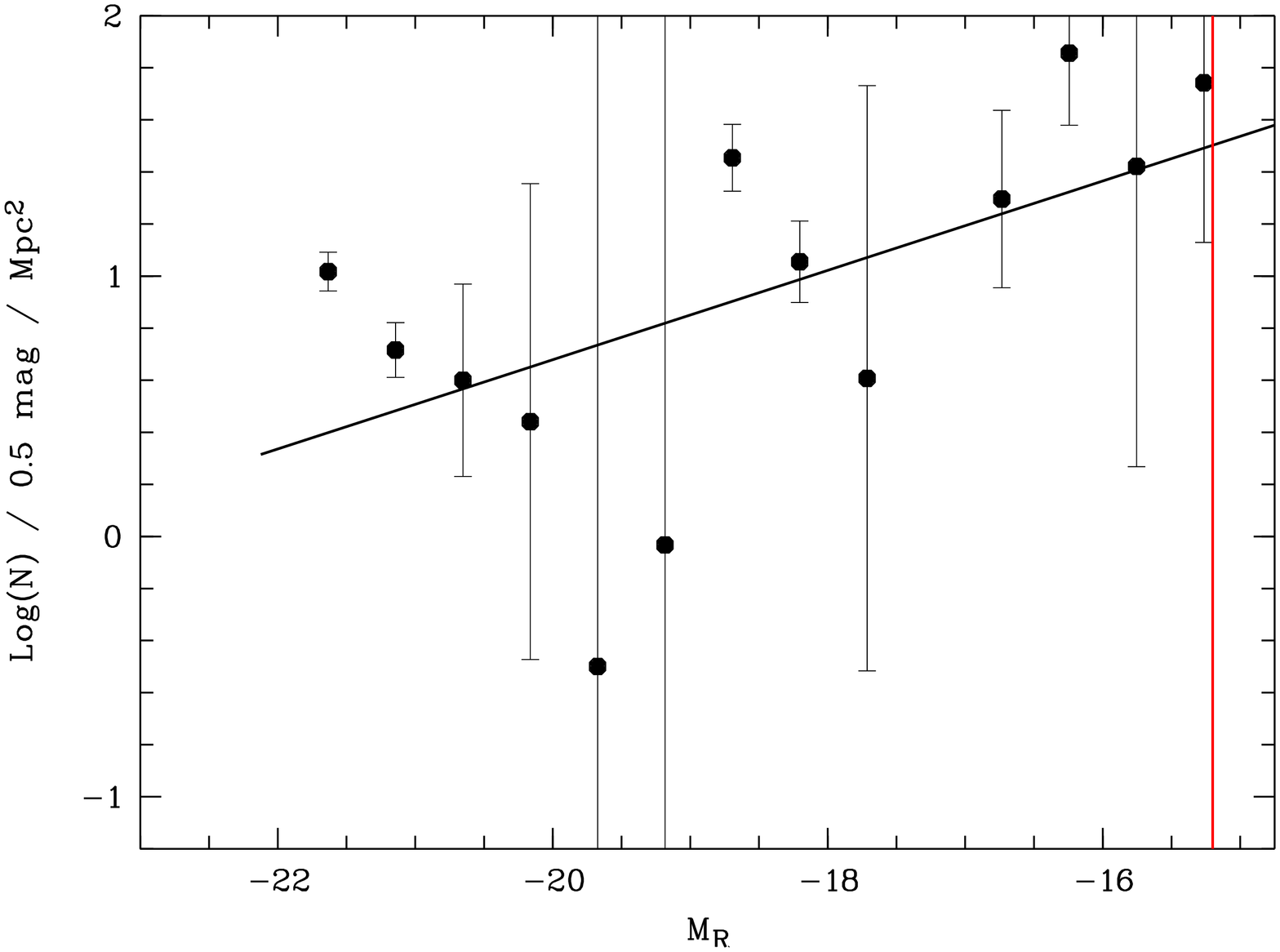,width=8cm,angle=270}}
\caption[]{Statistically-subtracted Rc-band galaxy luminosity
  functions of NGC~6034 (top), 1RXS J235814.4+150524 (middle), and RX
  J1119.7+2126 (bottom). The black lines show the mean luminosity
  functions in the M$_{Rc}$ intervals [-21, -18.]  (NGC~6034) and [-22.,
  -19.] (1RXS J235814.4+150524).
  The red lines are the adopted completeness levels. The blue crosses
  are from the spectroscopic catalogs.}
\label{fig:FDL2}
\end{figure}

Given the effects that are sometimes important for galaxy photometric type
selection on the membership criterion, we also computed luminosity
functions with other methods. We first used the spectroscopic catalog
for 1RXS J235814.4+150524 assuming that no selection bias was present
(this was not possible for RX J1119.7+2126 because of the few
spectroscopic members). For a given magnitude interval (chosen to
include at least ten galaxies with a spectroscopic redshift inside the
galaxy structure), we computed the ratio between the group
spectroscopic members and the field galaxies. We then applied this
ratio to the photometric catalogs. This allowed us to put a few
spectroscopic-based points in Fig.~\ref{fig:FDLzphot}. These points
are within the error bars of the photometric redshift-based galaxy
luminosity functions. However, spectroscopy is not always deep enough
to sample the faint magnitude regime.  We therefore also computed
luminosity functions with a pure statistical approach. We statistically 
estimated the back- and foreground contributions along the considered lines 
of sight with the comparison fields already used in Adami et
al. (2007b). Since they were observed in sky regions of low extinction
with the same telescope, CCD, and Rc filter, they allowed us to
statistically subtract field galaxies. Results are shown in
Fig.~\ref{fig:FDL2} for the Rc band. Despite the larger error bars,
results are consistent with Fig.~\ref{fig:FDLzphot}
except perhaps for RX J1119.7+2126.

The computed luminosity functions also allow us to compute the total R-band 
luminosity of the three groups. Simply integrating the photometric redshift-based 
galaxy luminosity functions in the magnitude ranges considered in 
Fig.~\ref{fig:FDLzphot} and assuming L$\odot$$_R$=4.62, we obtained the total 
R-band luminosities listed in Table~\ref{tab:caract}. It is 
interresting to note that RX J1119.7+2126 has a very low optical luminosity
which agrees well with the estimates of Jones et al. (2003). Moreover, we also 
see that NGC~6034 and 
1RXS J235814.4+150524 have similar R-band (and X-ray bolometric) luminosities, 
even though one is a normal group and the other is a fossil group. 
NGC~6034, however, is a group with galaxies spread over a wide range of magnitudes, 
as opposed to 1RXS J235814.4+150524, which shows the well known 2-magnitude gap 
after the group dominant galaxy, many relatively bright 
galaxies (M$_R$$\sim$-20) after this gap, and no galaxy members fainter than 
M$_R$$\leq$-19.

\begin{figure}
\centering \mbox{\psfig{figure=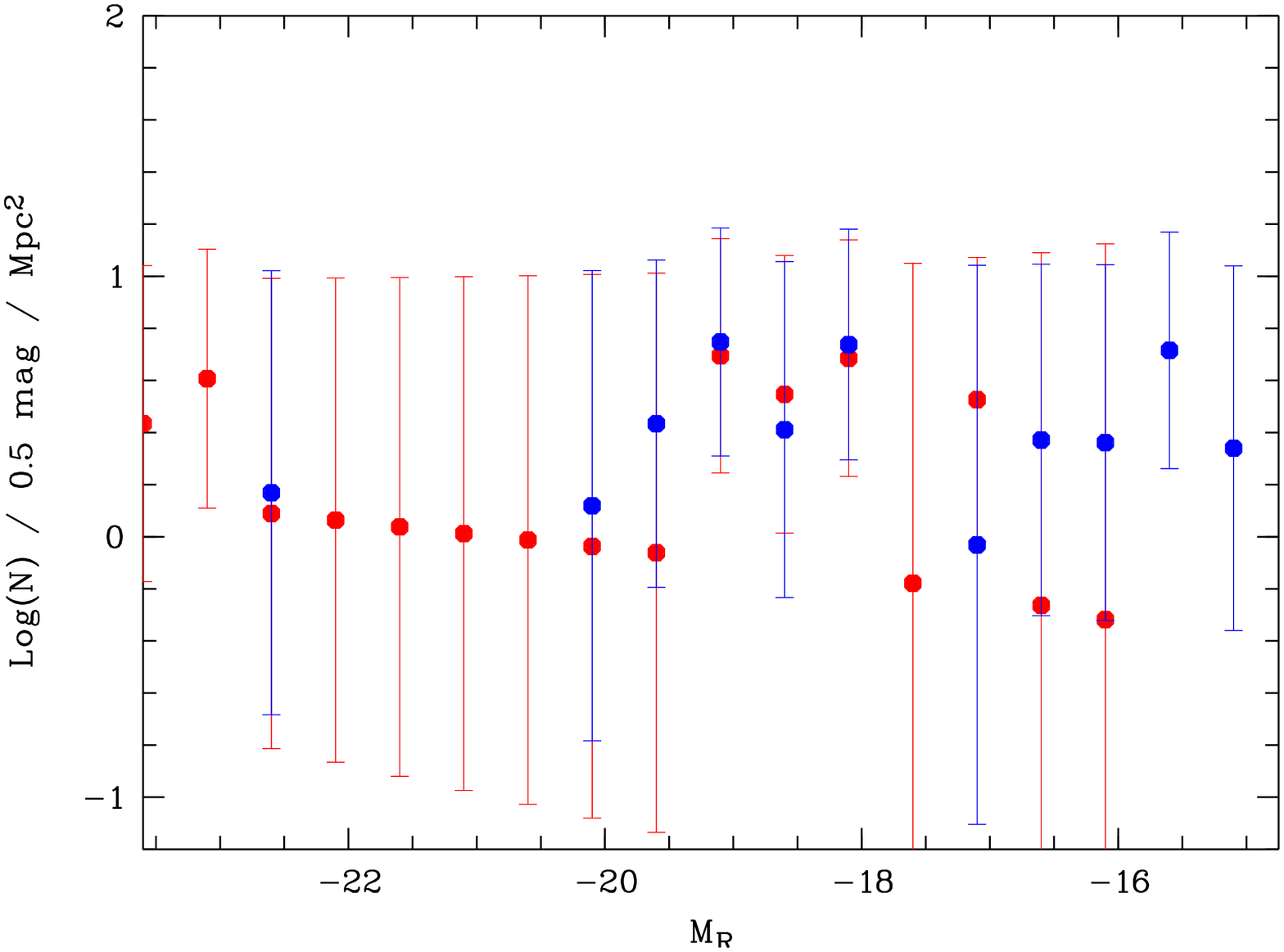,width=8cm,angle=270}}
\centering \mbox{\psfig{figure=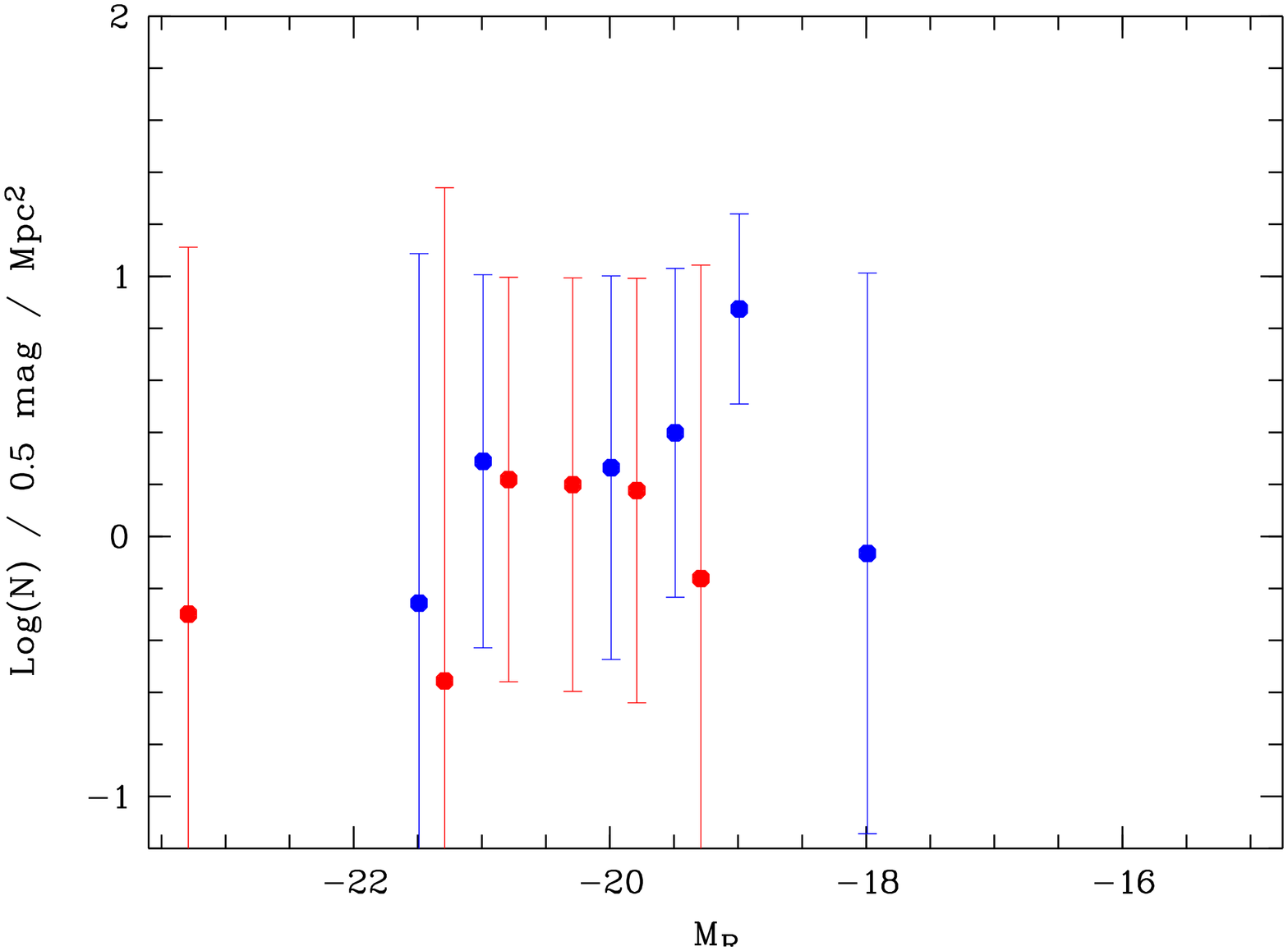,width=8cm,angle=270}}
\centering \mbox{\psfig{figure=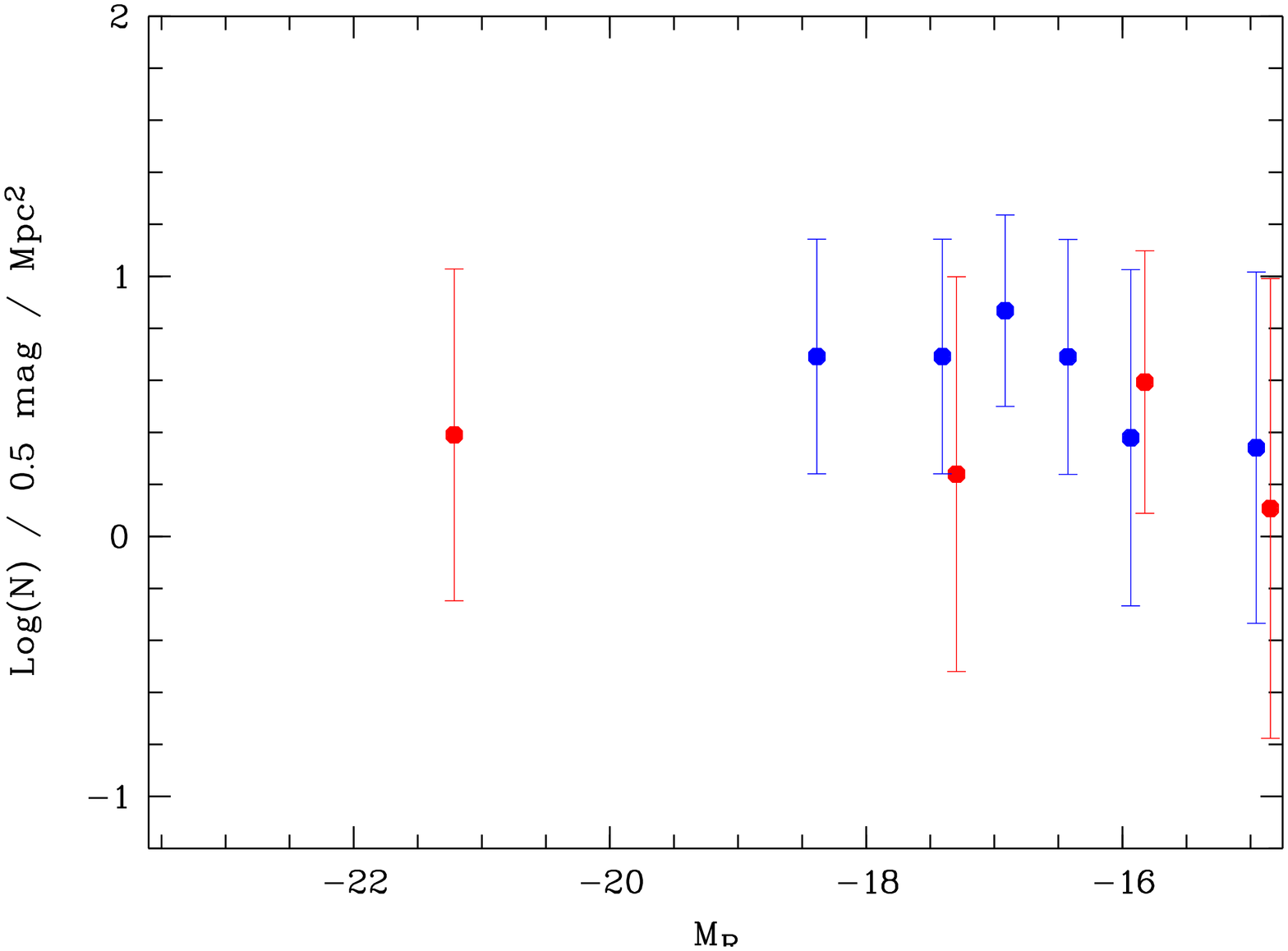,width=8cm,angle=270}}
\caption[]{Photometric redshift-based Rc band galaxy luminosity
  functions of NGC~6034 (top), 1RXS J235814.4+150524 (middle), and RX
  J1119.7+2126 (bottom). Red dots are for early-type or early spiral galaxies
  (types$\leq$34) and blue dots are for late spiral galaxies
  (types$\geq$35).}
\label{fig:FDLtype}
\end{figure}

The last question we adress in this section is the
relative contribution of the different galaxy photometric types to
these luminosity functions. For example, we showed that in massive
structures such as the Coma cluster, the faint parts of the luminosity
functions were mainly populated by late-type objects (see Adami et
al. 2007a). We detect a similar behavior for the groups (see 
Fig.~\ref{fig:FDLtype}). Keeping in mind that we are dealing with catalogs of 
potential structure members biased toward early-type galaxies and despite the 
less efficient selection of group members for late-type galaxies, these
galaxies are dominant compared to the early photometric types at the
faintest magnitudes. 

\subsection{Galaxy stellar population ages}

We confirmed in Adami et al. (2008) that short bursts of induced star
formation in galaxies that are just enter to the potential well of structures
were likely to occur in Coma. Since fossil groups are expected to
have typical X-ray gas contents, we should detect the same
behavior. We therefore computed these ages for the spectroscopic group
member galaxy samples. By fixing the redshift to the spectroscopic
value, we minimized the number of free parameters in the BC03 template
fitting.  Despite the limited samples,
Figs.~\ref{fig:SS6034}, \ref{fig:SS2359}, and \ref{fig:SS1119} seem
to show that the galaxies with recent stellar populations are
preferentially located in the group outskirts. 

For RX J1119.7+2126, there is one galaxy very
close to the dominant galaxy that also shows a relatively young
stellar population (4 Gyr versus 11 Gyr for the dominant galaxy).
This could be explained if this small galaxy already experienced close
encounters with the dominant galaxy, which could have induced bursts of star
formation. Another explanation would be that this galaxy is not dynamically
linked to the group and is just a young group-passing-through object, although 
its spectroscopic redshift (z=0.06134) is very similar to the group redshift 
(z=0.061).

\section{Summary}

By gathering imaging and spectroscopic data from various sources, we
have analysed the properties of three groups of galaxies.  Our main
results are the following:

\begin{itemize}

\item The large-scale environment of 1RXS J235814.4+150524 is quite
  poor and diffuse (even if its galaxy density map does show a
  clear signature of a surrounding cosmic web), and therefore cannot 
  provide many infalling galaxies.  RX J1119.7+2126
  appears to be very isolated, as previously shown by Adami et
  al. (2007b). On the other hand, Lopes de Oliveira et al. (2010) have
  shown that the cosmic environment of NGC~6034 is very rich.

\item At the group scale, 1RXS J235814.4+150524 shows no substructure
and seems to have a flatter density profile in its center than NGC6034.
  RX J1119.7+2126 do not show any galaxies present in the immediate
  vicinity of the dominant galaxy.

\item A red sequence is discernable for all three groups in a
  color-magnitude diagram. The luminosity functions derived with
  photometric redshift selection and with statistical background
  subtraction show comparable shapes, and also agree with the few
  points obtained based on spectroscopic redshifts. The two fossil groups
  show a dip after the dominant galaxy. Luminosity functions have a
  regular shape down to the completeness level for RX J1119.7+2126 and
  NGC~6034 with a nearly flat slope for the faintest magnitudes. There is clear 
  lack of faint galaxies for 1RXS J235814.4+150524 for the faintest
  magnitudes. This agrees well with the slightly decreasing 
  luminosity functions of Proctor et al. (2011) at M$_R$$\geq$-20 when 
  considering areas inside half the virial radius. The faint parts of the 
  luminosity functions of our two fossil groups finally appear dominantly 
  populated by late-type galaxies.

\item Galaxies with recent stellar populations seem preferentially
  located in the group outskirts. This is expected if bursts of star
  formation occured when these galaxies entered the groups.

\item RX J1119.7+2126 is definitely classified as a fossil group; 1RXS
  J235814.4+150524 also has properties very close to those of a fossil
  group, while we confirm that NGC~6034 is a normal group.

\end{itemize}

\begin{acknowledgements}
The authors thank the referee for useful and constructive comments.
  We are grateful to the organizers of the 6th Neon Observing
  School. We gratefully acknowledge the contributions of the students
  of the 2005/2008 classes of the Aix-Marseille I AER M2. We also
  thank the whole XMM-LSS team for their help.
\end{acknowledgements}

\appendix

\section{Spectroscopic redshifts}

\begin{figure}
\centering \mbox{\psfig{figure=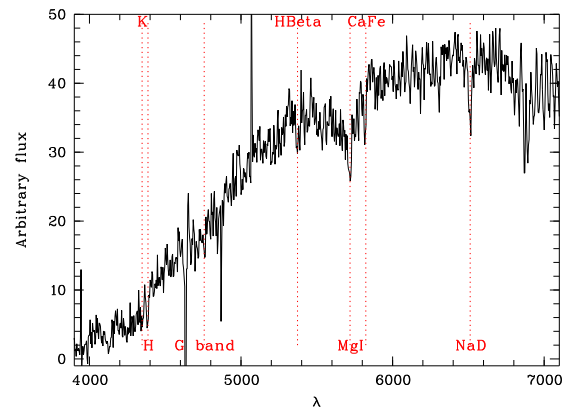,width=8cm,angle=270}}
\caption[]{Spectrum of 2MASX J23480439+1442557 observed with the B$\&$C 
spectrograph. The identified lines are shown in red.}
\label{fig:J23480439}
\end{figure}

This section gives the measured spectroscopic redshifts for the three
groups (Tables~\ref{tab:spectrofaint1119}, ~\ref{tab:spectroohp}, 
~\ref{tab:spectrofaint2359}, and ~\ref{tab:spectrofaint60341}) and a typical
spectrum for the bright galaxies of the wide study of the neighborhood of 
1RXS J235814.4+150524 (2MASX J23480439+1442557: Fig.~\ref{fig:J23480439}).

\begin{table*}
  \caption{Redshifts used for the deep study of the central area of RX J1119.7+2126. 
    Cols.~1 and 2: J2000.0 coordinates of the object (in decimal degrees); 
    col.~3: redshift (the number of significant digits depends on the spectrum quality, galaxies marked with a * are potential group members); 
    col.~4: nature of the object 
    (galaxy, active object, star);
    col.~5: source (TNG, NED).  The typical  uncertainty 
    on  non-literature redshifts is 0.001. Given the low signal-to-noise ratio of 
    the TNG spectra, we chose to use mainly literature redshifts 
    when available. We note that the group velocity dispersion was computed 
    without the z=0.0565 galaxy.}
\begin{tabular}{lllll}
\hline
\hline
RA & DEC & z & Nature & Source \\ 
\hline
 169.86131   & 	21.43701  & 0.362       & galaxy & TNG \\ 
 169.87387   & 	21.44891  & 0.261/0.261 & galaxy & TNG/TNG \\ 
 169.88421   & 	21.46373  & 0.2595	& galaxy & TNG \\ 
 169.88604   & 	21.38743  & 1.422       & active & TNG \\ 
 169.88833   &  21.46607  & 0.28100     & galaxy & NED  \\ 
 169.89172   & 	21.40683  & 0.3379	& galaxy & TNG \\ 
 169.90656   & 	21.43067  & 0.          & star & TNG \\ 
 169.90868   & 	21.44097  & 2.97        & active & TNG \\ 
 169.90989   & 	21.39666  & 0.3065      & galaxy & TNG \\ 
 169.91870   & 	21.43216  & 0.2755      & galaxy & TNG \\ 
 169.92206   & 	21.45909  & 0.361       & galaxy & TNG \\ 
 169.92260   & 	21.46334  & 0.          & star   & TNG \\ 
 169.92296   & 	21.42640  & 0.4345      & galaxy & TNG \\ 
 169.92513   &  21.43622  & 0.06134 *    & galaxy & NED \\ 
 169.92680   & 	21.45457  & 0.392       & galaxy & TNG \\ 
 169.93159   &  21.44756  & 0.06055/0.0603/0.0603 * & galaxy & NED/TNG/TNG  \\ 
 169.93549   &  21.46902  & 0.306       & galaxy & TNG \\ 
 169.93623   &  21.53648  & 0.14058     & galaxy & NED  \\ 
 169.93633   & 	21.51504  & 0.2275      & galaxy & TNG \\ 
 169.93824   &  21.51931  & 0.22931/0.228 & galaxy & NED/TNG \\ 
 169.94032   & 	21.50432  & 0.153         & galaxy & TNG \\ 
 169.94221   & 	21.42251  & 0.3975        & galaxy & TNG \\ 
 169.94622   &  21.48221  & 0.21409/0.2125/0.213 & galaxy & NED/TNG/TNG  \\ 
 169.96226   & 	21.51082  & 0.0565 *?       & galaxy & TNG \\ 
 169.96593   & 	21.51904  & 0.405	  & galaxy & TNG \\ 
 169.96682   &  21.50062  & 0.08247/0.082 & galaxy & NED/TNG  \\ 
 169.96832   & 	21.41800  & 0.1652	  & galaxy & TNG \\ 
 169.98881   & 	21.44085  & 0.059 *	  & galaxy & TNG \\ 
 169.99585   &  21.53157  & 0.05976 *       & galaxy & NED  \\ 
 169.99891   & 	21.45859  & 0.299	  & galaxy & TNG \\ 
 170.00296   & 	21.44578  & 0.2095/0.2104 & galaxy & TNG/TNG \\ 
 170.00600   & 	21.43058  & 0.429	  & galaxy & TNG \\ 
\hline
\end{tabular}
\label{tab:spectrofaint1119}
\end{table*}

\begin{table*}
  \caption{Redshifts used for the wide study of the neighborhood of 1RXS J235814.4+150524. Col.~1: NED galaxy name; col.~2: redshift; col.~3: SDSS 
    r' magnitude; col.~4: instrument (Carelec, Boller $\&$ Chivens or AFOSC); 
    col.~5: exposure time. The typical redshift 
    uncertainty for these galaxies is 0.0005 and the spectral resolution is 7~\AA/px.}
\begin{tabular}{lllll}
\hline
\hline
Name & z & r' & Ins. & Exp. time \\ 
 &  &  &  & (hours) \\ 
\hline
2MASX J23475868+1410367   & 0.0709 & 15.92 & Carelec  & 1   \\
2MASX J23480439+1442557   & 0.1052 & 16.16 & B$\&$C   & 1   \\
SDSS  J234826.30+160113.7 & 0.0012 & 14.17 & Carelec  & 1   \\
2MASX J23510162+1527106   & 0.2510 & 15.49 & AFOSC    & 1   \\
2MASX J23530977+1503278   & 0.0786 & 15.53 & B$\&$C   & 2.5 \\
2MASX J23543897+1604501   & 0.0755 & 16.06 & B$\&$C   & 2.5 \\
2MASX J23543994+1604561   & 0.0885 & 15.12 & B$\&$C   & 2.5 \\
2MASX J23551805+1531545   & 0.0747 & 15.98 & Carelec  & 1   \\
KUG   2358+159            & 0.0440 & 15.87 & Carelec  & 1   \\
2MASX J23582694+1513586   & 0.1527 & 16.04 & AFOSC    & 2   \\
2MASX J23592068+1440129   & 0.0928 & 16.06 & AFOSC    & 2   \\
2MASX J00001187+1405240   & 0.1035 & 15.96 & Carelec  & 1   \\
2MASX J00040323+1420083   & 0.0383 & 15.67 & Carelec  & 1   \\
SDSS  J000547.13+160838.1 & 0.1163 & 16.02 & Carelec  & 1   \\
2MASX J00061715+1354269   & 0.0757 & 16.18 & AFOSC    & 0.5 \\
2MASX J00075853+1510533   & 0.1131 & 16.00 & Carelec  & 1   \\
\hline
\end{tabular}
\label{tab:spectroohp}
\end{table*}

\begin{table}
  \caption{Redshifts used for the deep study of the central area of 1RXS J235814.4+150524.
    Cols.~1 and 2: coordinates of the object (decimal degrees); col.~3: redshift 
    (the number of significant digits depends on the precision achieved for the redshift
determination), galaxies marked with a * are potential group members); 
    col.~4: source    (TNG, NTT or NED).  The typical redshift uncertainty of the 
non-literature redshifts is 0.001. Given the good quality of the TNG 
spectra, we chose to use mainly these spectra.}
\begin{tabular}{llll}
\hline
\hline
RA & DEC & z & Source \\ 
\hline
 359.49886 &  15.10602 &	0.1527	       & TNG \\ 
 359.50628 &  15.11972 &	0.1778 *	       & TNG \\ 
 359.51304 &  15.11320 &	0.1786 *	       & TNG \\ 
 359.51911 &  15.14732 &	0.1783 *	       & TNG \\ 
 359.52384 &  15.14407 &	0.1553	       & TNG \\ 
 359.53523 &  15.12241 &	0.1471/0.15539 & TNG/NED \\ 
 359.54073 &  15.11647 &	0.0455	       & TNG \\ 
 359.54074 &  15.09299 &	0.1804 *	       & TNG \\ 
 359.54119 &  15.05579 &	0.0951/0.09537 & TNG/NED \\ 
 359.54428 &  15.07865 &	0.1748	       & TNG \\ 
 359.54479 &  15.10169 &	0.175	       & TNG \\ 
 359.55371 &  15.06013 &	0.1775/0.18900 * & TNG/NTT \\ 
 359.55694 &  15.09995 &	0.1795 *	       & TNG \\ 
 359.55717 &  15.10495 &	0.1288	       & TNG \\ 
 359.55917 &  15.11952 &	0.4105	       & TNG \\ 
 359.55965 &  15.14777 &	0.1895	       & TNG \\ 
 359.56001 &  15.09075 &	0.1769	       & TNG \\ 
 359.56027 &  15.15464 &	0.1531	       & TNG \\ 
 359.56263 &  15.04156 &	0.1783 *	       & TNG \\ 
 359.56280 &  15.09560 &	0.1786/0.1781 *  & TNG/NED \\ 
 359.56301 &  15.09571 &	0.1781/0.17843 * & TNG/NED \\ 
 359.56370 &  15.04343 &	0.1902	       & TNG \\ 
 359.56425 &  15.06989 &	0.3460	       & TNG \\ 
 359.56460 &  15.06430 &	0.1547/0.1535  & TNG/NED \\ 
 359.56483 &  15.06406 &	0.1535	       & TNG \\ 
 359.56519 &  15.08331 &	0.1891	       & TNG \\ 
 359.56843 &  15.13146 &	0.4937/0.4460  & TNG/NED \\ 
 359.56862 &  15.13926 &	0.1590	       & TNG \\ 
 359.56865 &  15.13907 &	0.1549	       & TNG \\ 
 359.56870 &  15.13147 &	0.446	       & TNG \\ 
 359.56910 &  15.09212 &	0.1794 *	       & TNG \\ 
 359.56978 &  15.05974 &	0.1783 *	       & TNG \\ 
 359.57040 &  15.10639 &	0.1755	       & TNG \\ 
 359.57131 &  15.09885 &	0.1745/0.1793  & TNG/NTT \\ 
 359.57407 &  15.15178 &	0.2961	       & TNG \\ 
 359.57923 &  15.08278 &	0.4938	       & TNG \\ 
 359.58374 &  15.10016 &	0.1783 *	       & TNG \\ 
 359.58437 &  15.13468 &	0.1785 *	       & TNG \\ 
 359.59139 &  15.10234 &	0.1824	       & TNG \\ 
 359.59736 &  15.11437 &	0.179 *	       & TNG \\ 
 359.60624 &  15.06930 &	0.1520	       & TNG \\ 
 359.61164 &  15.06756 &	0.1865	       & TNG \\ 
 359.62673 &  15.07429 &	0.1675	       & TNG \\ 
\hline
\end{tabular}
\label{tab:spectrofaint2359}
\end{table}

\begin{table}
  \caption{Redshifts used for the deep study of the central area of NGC~6034, taken 
    from the literature (SDSS). Cols.~1 and 2: coordinates of the object (decimal degrees); 
    col.~3: redshift, galaxies marked with a * are potential group members).}
\begin{tabular}{lll}
\hline
\hline
RA & DEC & z \\ 
\hline
240.75500 & 17.26940 & 0.111570 \\
240.76401 & 17.24150 & 0.134560 \\
240.76801 & 17.19090 & 0.036130 * \\
240.76801 & 17.24870 & 0.135860 \\
240.77200 & 17.19340 & 0.033120 * \\
240.77400 & 17.17230 & 0.033590 * \\
240.81200 & 17.23930 & 0.035130 * \\
240.82401 & 17.18380 & 0.034860 * \\
240.84399 & 17.05720 & 0.038910 \\
240.85300 & 17.28030 & 0.134290 \\
240.86600 & 17.27470 & 0.135400 \\
240.86700 & 17.19620 & 0.034520 * \\
240.86700 & 17.27220 & 0.134850 \\
240.87100 & 17.15840 & 0.035360 * \\
240.87900 & 17.18240 & 0.033770 * \\
240.88400 & 17.19840 & 0.033880 * \\
240.88800 & 17.16310 & 0.033180 * \\
240.89600 & 17.25580 & 0.034770 * \\
240.91400 & 17.18480 & 0.034080 * \\
240.91901 & 17.33800 & 0.042880 \\
240.92000 & 17.06900 & 0.137350 \\
240.92900 & 17.30370 & 0.096960 \\
240.93401 & 17.29740 & 0.035570 * \\
240.93600 & 17.28870 & 0.097110 \\
240.93700 & 17.13570 & 0.037940 \\
240.93700 & 17.23070 & 0.033110 * \\
240.95100 & 17.24060 & 0.036530 * \\
240.96100 & 17.23950 & 0.042600 \\
240.97701 & 17.07650 & 0.034850 * \\
240.98500 & 17.11100 & 0.034780 * \\
240.98599 & 17.30510 & 0.033390 * \\
240.99001 & 17.23890 & 0.139580 \\
241.00600 & 17.13920 & 0.099240 \\
241.01100 & 17.28240 & 0.033470 * \\
241.02299 & 17.22080 & 0.295020 \\
241.03000 & 17.23640 & 0.031450 \\
241.03200 & 17.20700 & 0.035030 * \\
241.04300 & 17.20510 & 0.034840 * \\
\hline
\end{tabular}
\label{tab:spectrofaint60341}
\end{table}


\begin{thebibliography}{}

\bibitem[]{} Adami  C., Nichol R.C., Mazure A., et al., 1998, A$\&$A 334, 765

\bibitem[]{} Adami  C., Durret F., Mazure A., et al., 2007a, A$\&$A 462, 411

\bibitem[]{} Adami  C., Russeil D., Durret F., 2007b, A$\&$A 467, 459

\bibitem[]{} Adami  C., Ilbert O., Pell\'o R., et al., 2008, A$\&$A 491, 681

\bibitem[]{} Adami  C., Mazure A., Pierre M., et al., 2011, A$\&$A 526, 18

\bibitem[]{} Arnouts  S., Cristiani S., Moscardini L., et al., 1999, MNRAS 310, 540

\bibitem[]{} Bertin E., Arnouts S., 1996, A$\&$A 117, 393

\bibitem[]{} Bertin E., Mellier Y., Radovich M., 2002, ASPC 281, 228

\bibitem[]{} Bertin E., 2006, ASPC 351, 112

\bibitem[]{} Bruzual G., Charlot S., 2003, MNRAS 344, 1000

\bibitem[]{} Carlberg R.G., Yee H.K.C., Ellingson E., et al., 1997, ApJ 485, L13

\bibitem[]{} Calzetti D., Heckman T. M., 1999, ApJ 519, 27

\bibitem[]{} Cousins A.W.J., 1973, Mem RAS 77, 223

\bibitem[]{} Cousins A.W.J., 1974, MNSSA 33, 149

\bibitem[]{} Cypriano E.S., Mendes de Oliveira C., Sodr\'e L. Jr., 2006, AJ 132, 514

\bibitem[]{} Dunkley J., Komatsu E., Nolta M.R., et al., 2009, ApJS 180, 306

\bibitem[]{} Garilli B., Fumana M., Franzetti P., et al., 2010, PASP 122, 827

\bibitem[]{} Guennou L., Adami C., Ulmer C., et al., 2010, A$\&$A 523, 21

\bibitem[]{} Hoyle F., Vogeley M.S., 2004, AlJ 607, 751

\bibitem[]{} Ilbert O., Tresse L., Zucca E., et al., 2005, A$\&$A 439, 863

\bibitem[]{} Ilbert O., Arnouts S., McCracken H.J., et al., 2006, A$\&$A 457, 841

\bibitem[]{} Jones L.R., Ponman T.J., Horton A., et al., 2003, MNRAS 343, 627

\bibitem[]{} Kodama T., Arimoto N., 1997, A$\&$A 320, 41

\bibitem[]{} Lopes de Oliveira R., Carrasco E.R., Mendes de Oliveira C., et al., 2010, AJ 139, 216

\bibitem[]{} Mulchaey  J.S., Zabludoff A.I., 1999, ApJ 514, 133 

\bibitem[]{} Proctor R.N., Mendes de Oliveira C., Dupke R., et al., 2011, MNRAS 418, 2054

\bibitem[]{} Santos W.A., Mendes de Oliveira C., Sodr\'e L.Jr., 2007, AJ 134, 1551

\bibitem[]{} Schneider D.P., Gunn J.E., Hoessel J.G., 1983, ApJ 264, 337

\bibitem[]{} Serna A., Gerbal D., 1996, A$\&$A 309, 65

\bibitem[]{} Thuan T.X., Gunn J.E., 1976, PASP 88, 543

\bibitem[]{} Ulmer M.P., Adami C., Covone G., et al., 2005, ApJ 624, 124

\bibitem[]{} Yoshioka T., Furuzawa A., Takahashi S., et al., 2004, Adv. Space Res. 34, 2525



\end{thebibliography}
\end{document}